\documentclass[12pt]{article}
\pdfoutput=1
\usepackage{lscape,epsfig,amsmath,setspace}

\usepackage[numbers]{natbib}
\usepackage{acronym}
\usepackage{enumerate}
\usepackage{color}

\usepackage{tikz}
\usetikzlibrary{shapes.geometric,arrows,positioning, backgrounds}
\tikzstyle{box} = [rectangle, rounded corners, minimum width=1.5cm, minimum height=0.5cm,text centered, draw=black]

\usepackage{amsmath,amsfonts,amssymb}
\usepackage[all]{xy}
\usepackage{psfrag}

\usepackage{fullpage} 
\usepackage{amsmath,amsfonts,amssymb}
\usepackage{psfrag}
\usepackage{epsfig}
\usepackage[all]{xy}

\usepackage[refpage]{nomencl}

\makenomenclature

\newcommand{\e}[1]{\ensuremath{{\rm E}[#1]}}
\newcommand{\ed}[2]{\ensuremath{{\rm E}_{#1}[#2]}}
\newcommand{\var}[1]{\ensuremath{{\rm Var}[#1]}}
\newcommand{\vard}[2]{\ensuremath{{\rm Var}_{#1}[#2]}}

\newcommand{\cov}[2]{\ensuremath{{\rm Cov}\left[#1,#2\right]}}
\newcommand{\covd}[3]{\ensuremath{{\rm Cov}_{#1}\left[#2,#3\right]}}

\newcommand{\covb}[2]{\ensuremath{{\rm Cov}[#1,#2]}}
\newcommand{\re}[2]{\ensuremath{{\rm r}_{#1}(#2)}\,}
\newcommand{\R}[3]{\ensuremath{{ R}_{#1}(#2,#3)}\,}

\newcommand{\be}{\begin{equation}}
\newcommand{\ee}{\end{equation}}
\newcommand{\ba}{\begin{eqnarray}}
\newcommand{\ea}{\end{eqnarray}}
\newcommand{\bi}{\begin{itemize}}
\newcommand{\ei}{\end{itemize}}
\newcommand{\bn}{\begin{enumerate}}
\newcommand{\en}{\end{enumerate}}
\newcommand{\bfi}{\begin{figure}}
\newcommand{\efi}{\end{figure}}

\newcommand{\mx}{\mathcal{X}}
\newcommand{\mk}{\mathcal{K}}
\newcommand{\ml}{\mathcal{L}}
\newcommand{\mm}{\mathcal{M}}

\newcommand{\qq}{\quad \quad}

\newcommand{\ian}[1]{\textcolor{black}{#1}}

\newcommand{\iv}[1]{\textcolor{black}{#1}}
\newcommand{\ivm}[1]{\textcolor{black}{#1}}

\newcommand{\seq}[2]{#1,\dots,#2}
\newcommand{\seqsub}[3]{\seq{#1_{#2}}{#1_{#3}}}
\newcommand{\seqsup}[3]{\seq{#1^{#2}}{#1^{#3}}}
\newcommand{\seqsupb}[3]{\seqsup{#1}{(#2)}{(#3)}}
\newcommand{\seqsupbf}[3]{\seq{f(#1^{(#2)})}{f(#1^{(#3)})}}

\newcommand{\shortdist}[1]{$ a^{#1} = x - x^{#1} $}

\newcommand{\covxB}[1]{\covb{f(x)}{f(x^{#1})}}

\newcommand{\covDB}[2]{\covd{#2}{f(x)}{#1}}
\newcommand{\covDff}[3]{\covd{#1}{f(#2)}{f(#3)}}

\newcommand{\ef}[2]{\ed{#1}{f(#2)}}

\newcommand{\rx}[3]{\re{#1}{#2-#3}}
\newcommand{\rxB}[2]{\rx{#1}{x}{x^#2}}
\newcommand{\ra}[1]{\re{J_#1}{a^#1}}
\newcommand{\raK}{\ra{K}}

\newcommand{\rap}[1]{\re{J_#1}{a'^#1}}
\newcommand{\rapK}{\rap{K}}

\newcommand{\rp}[1]{\rx{P^{#1}}{x}{x'}}

\newcommand{\raa}[1]{\re{J_{#1}}{a^#1-a'^#1}}
\newcommand{\rara}[1]{\re{J_{#1}}{a^{#1}}  \re{J_{#1}}{a'^{#1}}}
\newcommand{\raaK}{\raa{K}}

\newcommand{\Ra}[1]{\R{J_#1}{a^#1}{a'^#1}}
\newcommand{\RaK}{\Ra{K}}

\newcommand{\ys}{y^{(s)}}
\newcommand{\zs}{z^{(s)}}
\newcommand{\zsK}{z^{(s)K}}

\newcommand{\xK}{x^K}
\newcommand{\xL}{x^L}
\newcommand{\xKL}{x^{KL}}
\newcommand{\xLK}{x^{LK}}
\newcommand{\aK}{a^K}
\newcommand{\aL}{a^L}

\newcommand{\aLK}{a^{LK}}

\newcommand{\xpL}{x'^L}

\newcommand{\xpLK}{x'^{LK}}
\newcommand{\apK}{a'^K}
\newcommand{\apL}{a'^L}

\newcommand{\apLK}{a'^{LK}}
\newcommand{\akH}{a^{K_H}}
\newcommand{\apkH}{a'^{K_H}}
\newcommand{\akHm}{a^{K_{H-1}}}
\newcommand{\apkHm}{a'^{K_{H-1}}}
\newcommand{\akT}{a^{K_T}}

\newcommand{\trick}[1]{\cov{f(x^{#1})}{#1} \var{#1}^{-1} =\; (1,0, \cdots, 0)}

\newcommand{\butnot}{\backslash}
\newcommand{\sds}{\sigma^2 \,}
\newcommand{\capdots}{\cap \cdots \cap}

\newcommand{\AEq}{\nonumber \\ & = &}
\newcommand{\AEqN}{\\ & = &}
\newcommand{\NL}{\nonumber \\ & & \qq }
\newcommand{\Df}[1]{\Delta f(#1)}
\newcommand{\tim}{\; \times \;}

\newcommand{\Rxz}[2]{\R{J_{#1}}{x^{#2}-x^{#2#1}}{\zs - z^{(s)#1}}}
\newcommand{\rpxz}[2]{\rx{P^{#1}}{x^{#2}}{\zs}}
\newcommand{\rpKLxL}{\rpxz{K \cup L}{L}}
\newcommand{\rpKL}{\rpxz{K \cup L}{}}

\newcommand{\rpKoL}{\rp{K \cup L}}
\newcommand{\rLnK}[1]{\re{J_L \butnot J_K}{#1}}
\newcommand{\rKnL}[1]{\re{J_K \butnot J_L}{#1}}
\newcommand{\rKL}[1]{\re{J_K \cap J_L}{#1}}
\newcommand{\rKoL}[1]{\re{J_K \cup J_L}{#1}}
\newcommand{\rK}[1]{\re{J_K}{#1}}
\newcommand{\rL}[1]{\re{J_L}{#1}}
\newcommand{\RK}[2]{\R{J_K}{#1}{#2}}

\newcommand{\RKnL}[2]{\R{J_K \butnot J_L}{#1}{#2}}

\newcommand{\xKh}{x^{K_h}}
\newcommand{\xpKh}{x'^{K_h}}
\newcommand{\aKh}{a^{K_h}}
\newcommand{\apKh}{a'^{K_h}}

\newcommand{\mkH}{\mk_H}
\newcommand{\KH}{K_H}
\newcommand{\KHm}{K_{H-1}}
\newcommand{\rKT}{\re{J_T}{a^{K_T}}}
\newcommand{\RH}{\R{{\KH}}{x}{x'}}
\newcommand{\RHm}{\R{{\KHm}}{x}{x'}}
\newcommand{\rT}[1]{\re{J_T}{#1}}

\newcommand{\rHnT}[1]{\re{J_H \butnot J_T}{#1}}
\newcommand{\rHmnT}[1]{\re{J_{H-1} \butnot J_T}{#1}}
\newcommand{\rpH}[1]{\re{P^H}{#1}}
\newcommand{\rpHm}[1]{\re{P^{H-1}}{#1}}
\newcommand{\zsKHm}{z^{(s)K_{H-1}}}
\newcommand{\xKhKHm}{x^{K_hK_{H-1}}}
\newcommand{\xpKhKHm}{x'^{K_hK_{H-1}}}
\newcommand{\rTnh}[1]{\re{J_T \butnot J_h}{#1}}
\newcommand{\rTh}[1]{\re{J_T \cap J_h}{#1}}
\newcommand{\rHmh}[1]{\re{J_{H-1} \cap J_h}{#1}}

\newcommand{\rh}[1]{\re{J_h}{#1}}
\newcommand{\xKhKT}{x^{K_hK_T}}
\newcommand{\xKT}{x^{K_T}}
\newcommand{\aKhKT}{a^{K_hK_T}}

\newcommand{\Rg}[4]{\ensuremath{{ R}^{(#1)}_{#2}(#3,#4)}\,}
\newcommand{\gmm}{\gamma - 1}
\newcommand{\Kg}{K_\gamma}
\newcommand{\KG}{K_\Gamma}
\newcommand{\KGm}{K_{\Gamma - 1}}

\newcommand{\rgnGm}[1]{\re{J_\gamma \butnot J_{\Gamma - 1}}{#1}}
\newcommand{\aKg}{a^{K_\gamma}}
\newcommand{\apKg}{a'^{K_\gamma}}
\newcommand{\xKg}{x^{K_\gamma}}
\newcommand{\sumig}{\sum_{i=2}^\gamma}
\newcommand{\sumbG}{\sum_{b \subset \Gamma, \, b_1 < ... < b_i = \gamma}}
\newcommand{\sumihm}{\sum_{i=1}^{h-1}}
\newcommand{\sumihz}{\sum_{i=0}^{h-1}}
\newcommand{\sumTHm}{\sum_{T \subseteq H_{-1}, \, |T| = i}}
\newcommand{\sumTH}{\sum_{T \subseteq H, |T| = i}}
\newcommand{\prodli}{\prod_{l=1}^{i-1}}
\newcommand{\blm}{b_l - 1}
\newcommand{\KBlm}{K_{B_l - 1}}
\newcommand{\Kbl}{K_{b_l}}

\newcommand{\rblnBlm}[1]{\re{J_{b_l} \butnot J_{B_l - 1}}{#1}}
\newcommand{\Kblp}{K_{b_{l+1}}}
\newcommand{\xKb}{x^{K_b}}
\newcommand{\rHmnToh}[1]{\re{J_{H-1} \butnot (J_T \cup J_h)}{#1}}
\newcommand{\rhnHm}[1]{\re{J_h \butnot J_{H-1}}{#1}}
\newcommand{\rHmhnT}[1]{\re{(J_{H-1} \cap J_h) \butnot J_T}{#1}}
\newcommand{\rHmohnT}[1]{\re{(J_{H-1} \cup J_h) \butnot J_T}{#1}}
\newcommand{\rToh}[1]{\re{J_T \cup J_h}{#1}}

\newcommand{\ro}[1]{\re{1}{#1}}
\newcommand{\rtt}[1]{\re{\{2,3\}}{#1}}
\newcommand{\Rtt}[2]{\R{\{2,3\}}{#1}{#2}}
\newcommand{\aM}{a^M}
\newcommand{\xM}{x^M}

\newcommand{\ifelsearray}[4]{\left\{ \begin{array}{r@{\quad \mbox{if} \quad}l} #1 & #2 \\ #3 & #4  \end{array} \right.}
\def \simulator { f }

\def \expectation  {\textrm{E}}

\def \Sig { \mathbf{\Sigma} }

\def \cl { \theta }

\def \real { \mathbb{R} }

\def\spacingset#1{\renewcommand{\baselinestretch}%
{#1}\small\normalsize} \spacingset{1}

\begin{document}

\title{Efficient Emulation of Computer Models Utilising Multiple Known Boundaries \ian{of Differing Dimensions}}

\author{Samuel Jackson\footnote{s.e.jackson@soton.ac.uk}, Ian Vernon$^\dagger$ \\
  \hspace{1cm} \\
\small *Southampton Statistical Sciences Research Institute, University of Southampton,  \\
\small University Road, SO17 1BJ, Southampton, UK \\
\small ${ }^\dagger$Department of Mathematical Sciences, Durham University, \\
\small Stockton Road, DH1 3LE, Durham, UK
}

\maketitle

\normalsize

\abstract{Emulation has been successfully applied across a wide variety of scientific disciplines for efficiently analysing computationally intensive models. \iv{We develop known boundary emulation strategies} 
which utilise the fact that, for many computer models, there exist hyperplanes in the input parameter space for which the model output can be evaluated far more efficiently, whether this be analytically or just significantly faster using a more efficient and simpler numerical solver.  The information contained on these known hyperplanes, or boundaries, can be incorporated into the emulation process via analytical update, thus involving no additional computational cost. \ian{In this article, we show that such} analytical updates are available for multiple boundaries of various dimensions. We subsequently 
demonstrate which \iv{configurations} of boundaries such analytical updates are available for, in particular by presenting a set of conditions that such a set of boundaries must satisfy.  We demonstrate the powerful computational advantages of the known boundary emulation techniques developed on both an illustrative low-dimensional simulated example and a scientifically relevant and high-dimensional systems biology model of hormonal crosstalk in the roots of an Arabidopsis plant.}



\small

\section{Introduction \label{intro} }

Computer models, or simulators, have been used across a wide range of disciplines
to help understand the behavioural dynamics of physical systems.
Such computer models are often high-dimensional\ivm{, due to them possessing large numbers of input parameters,} and take a substantial amount of time to evaluate.
As a result,  performing a full uncertainty analysis of model behaviour - a critical part of any scientific study that requires model evaluations at a vast number of parameter combinations - may be unfeasible.
For this reason, emulators are frequently used as \iv{fast} statistical approximations to computer model output, providing a predicted value at any input and a corresponding measure of uncertainty, given that the model has been evaluated for a set of training inputs \citep{DACE89, Higdon04_prediction}.  
Emulation has been successfully applied across a variety of scientific disciplines, such as astrophysics \citep{Higdon04_prediction, GFBUA, EECECF}, climate science \citep{GFDEMEP, HMERCMPS, SECMP}, engineering \citep{GPESFEM}, \iv{epidemiology} \citep{BHMCIDM, ABCSBI}, and volcanology \citep{USCMQVH, PPGPE}.
Improved emulation strategies therefore have the potential to benefit many scientific areas, allowing more accurate analysis at lower computational cost.

\cite{KBECCM} describe an advance in emulation strategy that can lead to substantial improvements in emulator performance
by exploiting the fact that 
there often exist parameter settings for which computer model output can be evaluated far more efficiently (whether this be analytically or just significantly faster using a simpler numerical solver). 
This may be possible as a result of allowing various modules to decouple from more complex parts of the model,
particularly when certain parameters are set to zero.
Such parameter settings commonly lie across boundaries or hyperplanes of the input parameter space, 
\iv{hence leading to effectively known model behaviour on these boundaries that impose} constraints on the emulator itself \iv{(note that such Dirichlet boundary conditions on the emulator are distinct from the Dirichlet boundary conditions that could be imposed on the computer model \ivm{itself}).}
The information on these known boundaries can be incorporated into the emulation process via analytical update, thus involving no additional computational cost.
\iv{This is preferable to the approach explored by \cite{Tan:2017aa}, which uses substantial extra modelling and multiple extra emulator parameters (each requiring estimation) to ensure consistency with the known boundary. In contrast, the approach in \cite{KBECCM} includes no extra modelling, and zero additional parameters, instead updating the Gaussian Process (GP) style emulator with the boundary information in a natural way.}

In this article, we extend the work of the literature to show that such analytical updates are available for multiple boundaries of various dimensions.  In particular,  we 
demonstrate 
which \iv{configurations} of boundaries such analytical updates are available for. \ian{The results of this article both provide analytical insights and are directly applicable to the analysis of many realistic physical systems represented by computer models.} \iv{We demonstrate this by applying the methodology to a scientifically relevant model of hormonal crosstalk in the roots of Arabidopsis Thaliana. 
Due to the ease and substantial benefits of including known boundaries when emulating the Arabidopsis model, we would suggest that future Uncertainty Quantification (UQ) analyses of scientific models include a phase of identification and incorporation of known boundaries, if they are found to exist, as standard practice}.

The remainder of this article is organised as follows.  In Section \ref{1KB}, we review and extend the work of \cite{KBECCM} to the case of a single known boundary of any dimension (as opposed to $ p - 1 $, where $ p $ is the number of input components to the computer model).  Section \ref{MBVD} extends the theory to multiple boundaries of various dimensions,  
covering 
which \iv{configurations} of boundaries may be incorporated into an emulator analytically, before \iv{exhibiting} a low-dimensional illustrative example.  
Section \ref{AMAM} applies the emulation techniques to a current systems biology model of Arabidopsis Thaliana, with the article being concluded in Section \ref{conc}.


\section{Known Boundaries of Dimension $ p - k $ \label{1KB} }

This section reviews the work presented in \cite{KBECCM}, whilst extending it by allowing the known boundaries to be of any dimension.

\subsection{Emulation of Computer Models \label{1KB:ECM} }

We consider a computer model $ \simulator(x) $, where $ x \in \mx $ denotes a $p$-dimensional vector containing the computer model's input parameters, and $ \mx \subset \real^p $ is a pre-specified input parameter space of interest.  We assume that $ \simulator(x) $ is univariate, however, the results presented directly generalise to the corresponding multivariate case, with acceptable correlation structure, as discussed further in Appendix \ref{ME}.  We make the judgement, consistent with \ivm{much} of the computer model literature, that $ f(x) $ has a product correlation structure:
\begin{equation}
\covb{f(x)}{f(x')} \;=\; \sds r(x-x') \;=\; \sds \prod_{j=1}^p r_j(x_j - x'_j)  \label{eq_cor_struc}
\end{equation}
with $ r_j(0) = 1 $, corresponding to deterministic $ \simulator(x) $.  For example, a common choice is the Gaussian correlation function, given by
\begin{equation}
\cov{f(x)}{f(x')} = \sds \exp \left( - \sum_{j=1}^p \left\{ \frac{x_j - x_j^\prime}{\cl_j} \right\}^2 \right)  \label{GF}
\end{equation} 

If we perform a set of runs at locations $X_D=\{ x^{(1)},\dots,x^{(n)} \}$ over the input space of interest $\mx$, giving computer model outputs 
as the column vector $D = (f(x^{(1)}),\dots,f(x^{(n)}))^T$, then we can update our beliefs about the computer model $f(x)$ in light of $D$.
This can be done either using Bayes theorem (if, \iv{say}, $f(x)$ is assumed to be a GP) or using the Bayes linear update formulae
(which, following \cite{TP}, treats expectation as primitive and requires only a second order specification~\citep{BLA,BLS}):
\ba
\ed{D}{f(x)} &=& \e{f(x)} + \cov{f(x)}{D} \var{D} ^{-1}(D- \e{D}) \label{BLE1}\\
\vard{D}{f(x)} &=& \var{f(x)} - \cov{f(x)}{D} \var{D}^{-1}\cov{D}{f(x)} \label{BLE2} \\
\covd{D}{f(x)}{f(x')} &=& \cov{f(x)}{f(x')} - \cov{f(x)}{D} \var{D}^{-1} \cov{D}{f(x')} \label{BLE3_univariate}
\ea
where $\ed{D}{f(x)}$, $\vard{D}{f(x)}$ and $\covd{D}{f(x)}{f(x')}$ are the expectation, variance and covariance of $f(x)$ adjusted by $D$~\citep{BLA,BLS}.
\iv{Although we will work within the 
 Bayes linear formalism, the derived results will apply directly to the fully Bayesian case, were one willing to make the additional assumption of full normality that use of a GP entails. In that case, all Bayes linear adjusted quantities can be directly mapped to the corresponding posterior versions,} 
for example, $\ed{D}{f(x)} \rightarrow \e{f(x)|D}$ and $\vard{D}{f(x)} \rightarrow \var{f(x)|D}$.
 See \cite{BLA} and \cite{BLS} 
 for a further discussion of the Bayes linear approach \ivm{and its foundational motivation}. 

As discussed in \cite{KBECCM}, since the results \iv{of this article} rely on the product correlation structure of the emulator, \iv{extension} of these methods to more general emulator forms
requires further calculation.

\subsection{Known Boundary Emulation \label{subsec:1KB:KBE} }

Let $ \mk $ be a $ p-k $-dimensional hyperplane to which directions $ x_{J_K} = \seqsub{x}{j_1}{j_k} $ are normal,  where $ J_K = \{ \seqsub{j}{1}{k} \} \subset P = \{ \seq{1}{p} \} $.  The position of the hyperplane is defined by $ x_{J_K} = \alpha^K = (\seqsub{\alpha^K}{1}{k}) $ for some $k$-vector of constants $ \alpha^K $, with all other $ x_j $ being unconstrained.  We consider the situation where $ f $ is analytically solvable along $ \mk $ (that is, we \ivm{know} $ f(x) $ for any $ x \in \mk $). 
We wish to update our emulator, and hence our beliefs about $f(x)$ at input point $x \in \mx$, in light of $\mk$.
We capture model behaviour along $ \mk $ by evaluating $ K = ( \seqsupbf{y}{1}{m} ) $ for a large \iv{but} finite number $ m $ of points on $ \mk $, denoted $ \seqsupb{y}{1}{m} $, \iv{however we structure our calculations so that they can be easily generalised to the case of continuous model evaluations on $\mk$, as shown in  \cite{KBECCM}}.
\iv{Naively plugging} these $m$ runs into the Bayes Linear update Equations (\ref{BLE1}), (\ref{BLE2}) and (\ref{BLE3_univariate}) by replacing $D$ with $K$ may be infeasible due to the size of the $m\times m$ matrix inversion $\var{K}^{-1}$ ($ m $ may need to be extremely large to capture all the information available from $ \mk $). 
A direct update of the emulator 
is therefore non-trivial, hence 
we show from first principles that this update can be performed analytically for a wide class of emulators.  This is done by exploiting a sufficiency argument briefly described in the supplementary material of \cite{BCCM} though, \iv{to our knowledge,} only utilised for the first time in the context of known boundary emulation in \cite{KBECCM}.

We begin by evaluating $ f(x^K) $, where $ x^K $ is the orthogonal
projection of $x$ onto the boundary $\mk$, and extending the collection of boundary evaluations, $K$, to be the $m+1$ column vector
$ K=(f(x^K),f(y^{(1)}),\dots,f(y^{(m)}))^T $.
Note that $ x_{j_i}^K = \alpha_i^K $ for $ j \in J_K $.  Crucially, we have that

\be
\trick{K} \label{eq_covvar}
\ee
arising from the first row of the somewhat trivial equation
$ \var{K} \var{K}^{-1} =\; \boldsymbol{I}_{(m+1)}  $,
where $\boldsymbol{I}_{(m+1)}$ is the identity matrix of dimension $(m+1)$.  

Equation~(\ref{eq_covvar}) 
is of particular value when considering the behaviour of $ f $ at the point of interest $x$.
As we have defined $x^K$ as the orthogonal projection of $x$ onto $\mk$, we can 
define \shortdist{K} to be the $p$-vector of shortest distance from  boundary $ \mk $ to $ x $. Note that the elements of $ a^K $ have the property that:
\be
a_j^K = \ifelsearray{x_j - x_j^K}{j \in J_K}{0}{j \in P^K = P \butnot J_K} \nonumber
\ee
where we define (for \ivm{two sets} $ A, B $) $ A \butnot B $ to be the elements in $A$ but not $B$. 
We also define:
\be 
r_J(q) = \prod_{j \in J} r_j(q_j)  \label{r_note}
\ee 
for a generic collection of indices $ J $ and vector of constants $ q $\ivm{. By partitioning the dimension indices $P= \{ \seq{1}{p} \}$ into 
$P=\{J_K,P^K\}$ we obtain the} following covariance expressions:
\ivm{
\begin{alignat}{5}
\covxB{K}  \;&= &\; \sds \rxB{P}{K} \;&= &&\; \;    \sds \raK   \re{P^K}{0}  \;=\;  \sds \raK \label{eq_shortcov2} \\  
\covb{f(x)}{f(\ys)}  \;&= &\; \sds \rx{P}{x}{\ys} \;&= &&\;\;  \sds \raK \,   \rx{P^K}{x^K}{\ys}  \nonumber \\
&& &= && \; \; \raK \, \covb{f(x^K)}{f(\ys)} \label{eq_prod_ra_xy} 
\end{alignat}  
since $ r_j(0) = 1$,} components $ J_K $ of $x^K$ and $\ys$ must be equal as they all lie on $\mk$ (that is, $x^K_j = \ys_j$ for $ j \in J_K $), \ivm{and $x_j-\ys_j = x_j^K - \ys_j$ for $ j \in P^K$}.  \ivm{From Equations \eqref{eq_shortcov2} and \eqref{eq_prod_ra_xy}}, the covariance between \ivm{$f(x)$} and the set of boundary evaluations \ivm{$K$} is given by
\begin{equation}
\cov{f(x)}{K}  = \raK \, \cov{f(x^K)}{K}     \label{eq_covfxKpre}
\ee
Using Equations~(\ref{eq_covvar}) and (\ref{eq_covfxKpre}) we obtain the important result that:
\begin{equation}
\cov{f(x)}{K} \var{K}^{-1} \;=\; \raK (1,0, \cdots, 0) \label{eq_covfxK}
\ee
As we have avoided the need to explicitly evaluate the intractable matrix inverse $\var{K}^{-1}$, we can find the Bayes Linear adjusted expectation for $f(x)$ with respect to $K$ analytically by combining Equations~(\ref{BLE1}) and (\ref{eq_covfxK}):
\ba
\ed{K}{f(x)} &=& \e{f(x)} + \raK (1,0, \cdots, 0) (K- \e{K}) \nonumber \\
&=&  \e{f(x)} + \raK \Delta f(x^K)  \label{eq_EK1}
\ea
where we have defined $\Delta f(\cdot) = f(\cdot) - \e{f(\cdot)}$.
We have thus eliminated the need to explicitly invert the large matrix $\var{K}$ entirely by exploiting the symmetric product correlation structure and Equation \eqref{eq_covvar}. 
Similarly, we find the adjusted covariance between $ f(x) $ and $ f(x') $ given the boundary $ \mk $, where $ f(x') $ is \ivm{the} model output at a second point $ x' $, using Equations \eqref{BLE3_univariate} and \eqref{eq_covfxK}, \ivm{and again exploiting the partition $P=\{J_K,P^K\}$:} 
\ba
\covd{K}{f(x)}{f(x')}  \!&=&\!\! \cov{f(x)}{f(x')} -  \raK (1,0, \cdots, 0) \cov{K}{f(x')} \nonumber \\
\!&=&\!\! \cov{f(x)}{f(x')} -  \raK  \cov{f(x^K)}{f(x'^K)} \rapK    \nonumber  \\
\!&=&\!\! \ivm{\sds \raaK \rp{K} \!-\!  \sds \raK \re{J_K}{0}  \rp{K}  \rapK }   \nonumber  \\
\!&=&\!\! \sds \RaK \, \rp{K}  \label{eq_covK2}
\ea
where the `updated correlation component' in the $x_{J_K}$ directions is given as
\be
\RaK \;=\; \raaK -   \raK \rapK   \label{R_note}
\ee
By setting $ x = x' $, we obtain an expression for the adjusted variance of $ f(x) $:
\be
\vard{K}{f(x)} = \sds(1 - \raK^2) \label{eq_VK1}
\ee
Equations~(\ref{eq_EK1}) and (\ref{eq_VK1}) give the expectation and variance of the emulator at a point $x$, updated by a known boundary $\mk$. As they require only evaluations of the analytic boundary function and the correlation function they can be implemented with trivial computational cost in comparison to a direct update by $K$. 
Useful insights into the sufficiency, stationarity and limiting behaviour, along with a generalisation of the above to continuous boundary evaluations $ K $, are discussed in \cite{KBECCM}.

\subsection{Three-dimensional Example \label{TE} }

\ivm{For illustration, we consider the problem of emulating the three-dimensional function:}
\begin{equation}
f(x) = \sin \left( \frac{x_1}{\exp(x_2)} \right) + \cos(x_3)   \label{TEq}
\end{equation}
over an input domain of interest given by $ [-2 \pi, 2 \pi ] \times [ - \pi/4, \pi/4] \times [-2 \pi, 2 \pi ]  $.  This simulated example will, throughout this article, take a prior expectation $ \expectation[f(x)] = 0 $, and a product Gaussian covariance structure, as given by Equation (\ref{GF}), with correlation length parameters $ \theta = (\pi, \pi/8, \pi) $ and variance parameter $ \sds = 2 $.  These values are adequate for the example presented, as illustrated in the diagnostic panels of the figures that follow.  Having said this, it is important and informative to explore the effect of varying the correlation length parameters on emulator predictions, particularly in combination with known boundaries.  We explore varying these parameters for the Arabidopsis model application in Section \ref{AMAM}.

We begin by assuming a known boundary $ \mk $ at $ (x_2, x_3) = (0, 0) $, hence that we can evaluate
$ f(x^K) = \sin(x_1) \, \ian{+\,1} $
for any point on the boundary \ian{$x^K \in  \mk$}.  We hence apply the \ivm{emulator expectation and variance update given by} Equations (\ref{eq_EK1}) and (\ref{eq_VK1}).  

\begin{figure}
\begin{center}
\includegraphics[width=13.5cm]{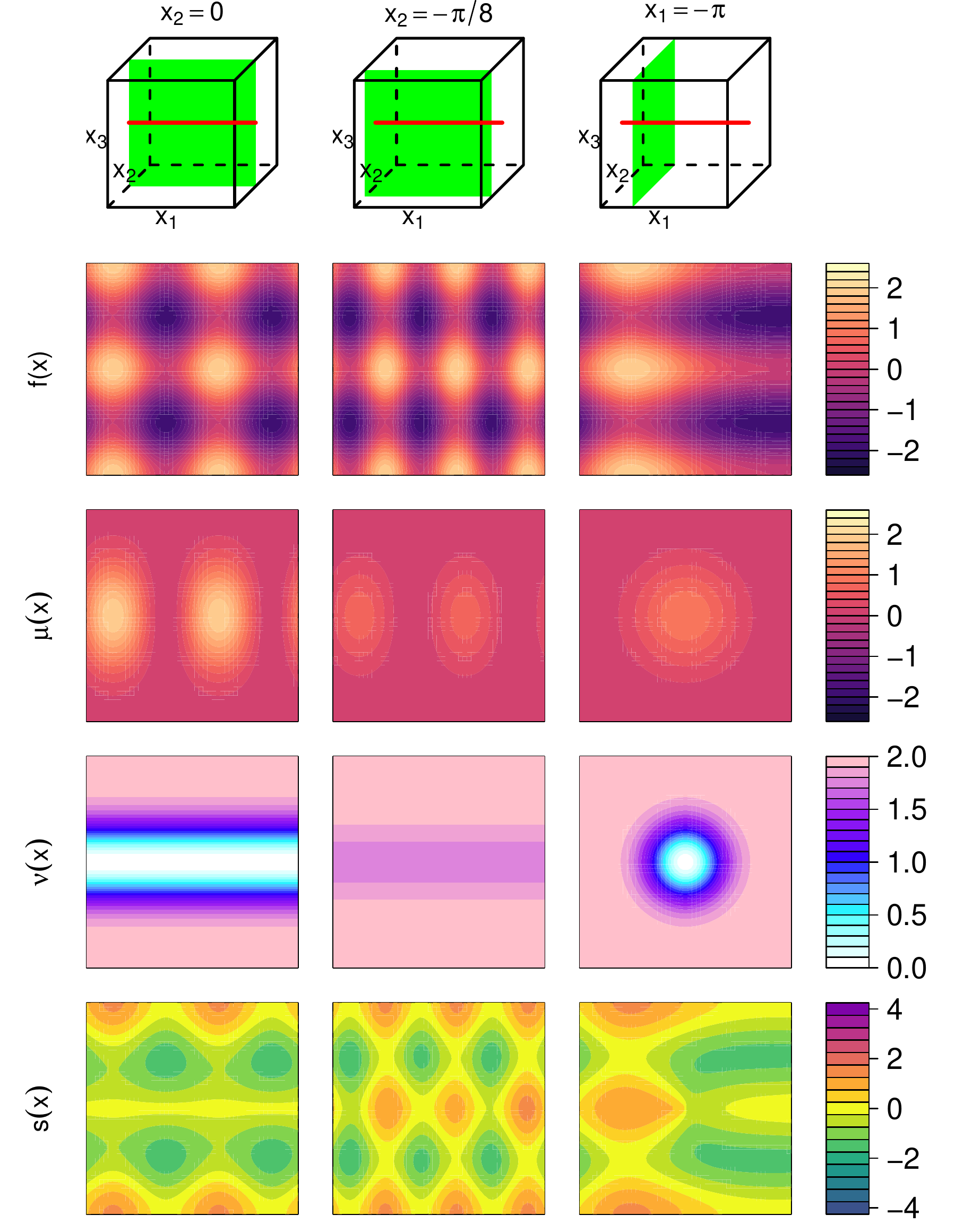}
\end{center}
\spacingset{1}
\caption{{Updating the emulator for the three-dimensional function given by Equation (\ref{TEq}) by a single boundary $ \mk $.  Rows from top to bottom, show: 1) position of known boundary (red line) and two-dimensional slices (green planes) over which the remaining plots are shown, 2) simulator function \ian{$f(x)$}, 3) emulator expectation \ian{$\mu(x)$}, 4) emulator variance \ian{$\nu(x)$}, 5) standardised errors \ian{$s(x)$}.  \ian{Columns from left to right show results on the three planes $ x_2 = 0 $, $ x_2 = -\pi/8 $ \ian{and} $ x_1 = -\pi $ respectively}.  Note that for each two dimensional plot, the variable with smaller index is along the horizontal axis.} \label{1BEx} }
\end{figure}

In order to illustrate the effect of the known boundary on the emulator, we examine emulator behaviour across two-dimensional slices (keeping one variable fixed) of the three-dimensional input space, as shown in Figure \ref{1BEx}.  The top row depicts the input space as a cube, with the one-dimensional boundary being illustrated by the red line.  The green planes are two-dimensional slices of the input space over which emulator and simulator behaviour are compared in the remaining plots.  The remaining rows show (from top to bottom) \iv{simulator} $ f(x) $ (for comparison purposes), emulator expectation $ \mu(x) = \ed{K}{f(x)} $, emulator variance $ \nu(x) = \vard{K}{f(x)} $ and standardised \ian{diagnostic} values $ s(x) = ( \ed{K}{f(x)} - f(x) ) / \sqrt{\vard{K}{f(x)} } $. 
In each case, the variable with smaller index is along the horizontal axis.  The left column of the figure shows the results for $ x_2 = 0 $.  Since this slice contains the known boundary, we see that for $ x_3 = 0 $ emulator expectation \ian{precisely} matches the true simulator function, and the variance goes to zero.  As we move further away from the boundary in the $ x_3 $ direction, the variance \ian{increases}.  Note that, since $ \mk $ is parallel to the $ x_1 $ direction, altering the value of this variable doesn't alter \ivm{the} emulator variance.  The middle column shows a slice away from the boundary ($ x_2 = -\pi/8 $).  Again, the smallest variance is at $ x_3 = 0 $, however, now it is not zero.  The right column shows $ x_1 = -\pi $.  In this case, the function is only known at the centre point $ (x_2, x_3) = (0,0) $ with variance increasing radially away from this point.  The diagnostic plots provide evidence for the validity of the emulator, with few parts of the input space having standardised errors greater than 2.

\subsection{Updating By Further Model Evaluations  \label{subsec:UFME} }

Since we have analytic expressions for $\ed{K}{f(x)}$, $\vard{K}{f(x)}$ and $\covd{K}{f(x)}{f(x')}$ we are now able to include additional computer model evaluations into the emulation process. 
To do this, we perform $n$ (expensive) evaluations of the computer model across $\mx$ to obtain $ D = ( f(x^{(1)}),...,f(x^{(n)}) ) $, and use these to supplement the evaluations, $K$, available on the boundary. We want to update the emulator by the union of the evaluations $D$ and $K$, that is to find $\ed{K \cup D}{f(x)}$, $\vard{K \cup D}{f(x)}$ and $\covd{K \cup D}{f(x)}{f(x')}$. This can be achieved via a sequential Bayes Linear update:

\small
\ba
 \ed{K \cup D}{f(x)}  &=&  \ed{K}{f(x)} + \covd{K}{f(x)}{D} \vard{K}{D}^{-1}(D- \ed{K}{D}) \qq \label{eq_BLmDK}\\
\vard{K \cup D}{f(x)}  &=&  \vard{K}{f(x)} - \covd{K}{f(x)}{D} \vard{K}{D}^{-1}\covd{K}{D}{f(x)}  \qq \label{eq_BLvDK}\\
\covd{K \cup D}{f(x)}{f(x')}  &=&  \covd{K}{f(x)}{f(x')} - \covd{K}{f(x)}{D} \vard{K}{D}^{-1} \covd{K}{D} {f(x')} \qq \label{eq_BLcDK} \\[-1.2cm] && \nonumber 
\ea
\normalsize where we first update our emulator analytically by $K$, and subsequently update these quantities 
by the evaluations $D$ \citep{BLS}. 
$n$ is typically of small/modest size due to the relative expense of evaluating the computer model, hence these calculations (in particular $\vard{K}{D}^{-1}$) will remain tractable. 

It is worth noting that users of \iv{standard} black-box \iv{GP} emulation packages may be \iv{unable to implement directly} the formulae of Equations~(\ref{eq_EK1}) and (\ref{eq_VK1}).  \iv{However, we see that due to underlying sufficiency arguments,} such a user can \iv{simply} add the $(n+1)$ trivial \iv{boundary} evaluations $ f(x^K) $ and $ D^K = ( f(x^{(1)K}),...,f(x^{(n)K}) ) $ to $ D $ to give $ D^* = \{ D, D^K, f(x^K) \} $, and then their black box Gaussian process package will produce results that 
precisely match Equations~(\ref{eq_BLmDK})-(\ref{eq_BLcDK}).  
\iv{However, this does require inverting} a potentially very large matrix for each collection of $ n' $ points to be emulated.  In comparison, \iv{directly} using the above analytic results 
\ian{(Equations~(\ref{eq_BLmDK})-(\ref{eq_BLcDK}) combined with~(\ref{eq_EK1}), (\ref{eq_covK2}), and (\ref{eq_VK1}))}
only requires inversion of a single $ n \times n $ matrix, regardless of the size of $ n' $.
For further discussion of this issue and its exacerbation for multiple known boundaries, see Appendix~\ref{KBBB}.


\section{Multiple Boundaries of Various Dimensions \label{MBVD} }

In this section, we begin by discussing the requirements for being able to analytically update by a second known boundary, before considering larger numbers of boundaries.

\subsection{Two Known Boundaries \label{ssec_two_KBs}}

Given the results of Section \ref{1KB}, we now proceed to consider analytical updating by a second known boundary $ \ml $ of dimension $ p - l $, orthogonal to the $ x_{J_L} = \seqsub{x}{j'_1}{j'_l} $ directions and \ivm{with} location defined by $ x_{J_L} = \alpha^L $.   As this section progresses, we will restrict the \ivm{form} of $ \ml $, for example, to being orthogonal or parallel to $ \mk $, however, for now we consider the general case.  
We define
$ L\;=\;\left(f(x^L),f(z^{(1)}),\dots,f(z^{(m)})\right)^T $ 
to be a vector of model evaluations, where $z^{(1)},\cdots,z^{(m)}$ constitute a large but finite number $ m $ of points along $ \ml $, and denote the $ p $-vector of shortest distance to $ x $ from its orthogonal projection $ x^L $ as \shortdist{L}.  We also define $ x^{LK} $ to be the sequential orthogonal projection of $ x $ first onto $ \ml $ and then onto $ \mk $, and correspondingly \shortdist{LK} to be the $p$-vector of shortest distance to $ x $ from this sequential projection.
In \cite{KBECCM}, it was demonstrated that analytic updating of two perpendicular or parallel $ p-1 $-dimensional boundaries $ \mk, \ml $ could be achieved.  As an example, the update for two perpendicular $p-1$-boundaries  was given by:

\small
\ba
\ed{K \cup L}{f(x)} &=& \e{f(x)} + \re{1}{a^K} \Delta f(x^K)  +  \re{2}{a^L} \Delta f(x^L) -  \re{1}{a^K} \re{2}{a^L} \Delta f(x^{LK}) \nonumber \\
\covd{K \cup L}{f(x)}{f(x')} &=& \sds \R{1}{a^K}{a'^K} \, \R{2}{a^L}{a'^L} \, \prod_{j=3}^{p} \re{j}{x - x'} \label{eq_covKL1eg} \\[-0.8cm] && \nonumber 
\ea
\normalsize
where it is assumed that $ \mk $ \ivm{and $\ml$ are} defined by $ x_1 = \alpha^K $ and $ x_2 = \alpha^L $, and we have utilised the notation of Equations \eqref{r_note} and \eqref{R_note}.  
The inclusion-exclusion nature of this result will also feature in the general results that follow, both in this section and in Section \ref{ssub_MKBs}.

More generally, to permit analytic updating by a second boundary $ \ml $ of dimension $ p-l $, as defined above, we need to find an analogous version of Equation (\ref{eq_covfxKpre}), which relates $ \covDB{L}{K} $ to $ \covd{K}{f(x^L)}{L} $.  
We begin by examining the expression for $ \covDff{K}{x^L}{\zs} $ in light of Equation (\ref{eq_covK2}), exploiting the notation of Equations \eqref{r_note} and \eqref{R_note}, 
\ivm{and using the partition $P^K = \{  J_L \butnot J_K,  P^{K \cup L}  \}$ where $ P^{K \cup L} = P \butnot (J_K \cup J_L) $:}
\small
\ba
\lefteqn{\covDff{K}{x^L}{\zs}} 
\AEq \sds \Rxz{K}{L} \rpxz{K}{L}
\AEq           \sds \rpKLxL \rLnK{x^L - \zs} 
                   \bigg( \rK{x^L - \zs}  -   \rK{x^L - x^{LK}}    \rK{\zs - \zsK} \bigg)
\AEq  \sds \rpKL  \bigg( \rKnL{x-\zs} - \, \rKnL{\aK}   \rKnL{\zs-\zsK}    \rKL{LK}^2 \bigg)   \label{2perpxLzs}
\ea
\normalsize
since $ \xL_j - \zs_j = 0 $ for $ j \in J_L $, $ \re{P^{K \cup L}}{\xL_j - \zs_j} = \re{P^{K \cup L}}{x_j - \zs_j} $ and $ \xL_j - \xLK_j = LK_j $ for $ j \in J_L $, where $ LK_j $ is a constant giving the orthogonal distance from $ \ml $ to $ \mk $ in the $ x_j $-direction.  In addition, note that we define throughout $ r_\emptyset(\cdot) = 1 $.  We then examine:
\small
\ba
\lefteqn{\covDff{K}{x}{\zs}} 
\AEq \sds \Rxz{K}{} \rpxz{K}{}
\AEq   \sds   \rpKL     \rLnK{x-\zs} 
            \bigg( \rK{x-\zs}    - \, \rK{x-\xK}   \rK{\zs - \zsK}   \bigg)
\AEq      \sds    \rpKL   \rLnK{\aL}  
           \bigg(   \rKnL{x-\zs}   \rKL{\aL}   
           \NL   - \,   \rKnL{\aK}  \rKL{\aK}  \rKnL{\zs-\zsK}    \rKL{LK}  \bigg)   \label{2perpxzs}
\ea
\normalsize
By combining Equations (\ref{2perpxLzs}) and (\ref{2perpxzs}) we obtain:
\small
\ba
\lefteqn{\covDff{K}{x}{\zs}} 
\AEq   \frac{ \rKnL{x-\zs}   \rKL{\aL}  - \,  \rKnL{\aK}  \rKL{\aK}  \rKnL{\zs-\zsK}    \rKL{LK} }{\rKnL{x-\zs}  - \rKnL{\aK}    \rKnL{\zs-\zsK}    \rKL{LK}\!^2}
\NL   \times \, \rLnK{\aL}   \covDff{K}{x^L}{\zs}       \label{2perpprod}
\ea
\normalsize
In order to obtain an equation analogous to Equation (\ref{eq_covfxKpre}) we need to be able to write $ \covDff{K}{x}{\zs} $ as a product of $ \covDff{K}{x^L}{\zs} $ and a function that does not depend on $ \zs $, thus permitting replacement of $ f(\zs) $ by $ L $ in the $ \covd{K}{\, \cdot \,}{f(\zs)} $ terms.  In general, this is not possible for a second boundary, since the required product correlation structure no longer exists, thus resulting in the appearance of $ \zs $ several times in the quotient in the expression on the right hand side of Equation (\ref{2perpprod}).   However, there are two general and commonly occurring cases when this dependency does not exist and our methods permit further analytic update by a second known boundary (and indeed further known boundaries, as discussed in Sections \ref{ssub_MKBs} and \ref{PerpSetsPar}). The first case is if we wish to update by a known boundary which is a hyperplane which is orthogonal to and intersecting the first, and the second case is if we wish to update by a known boundary that is a hyperplane which is parallel to the first, or a subplane thereof.  We discuss these two cases in the following sections.


\subsubsection{Two Intersecting Orthogonal Known Boundaries \label{ssec_two_orth_bound}}

If $ \mk $ and $ \ml $ are two intersecting orthogonal boundaries, that is $ \mk \cap \ml \neq \emptyset $, then the elements of $ \alpha^K $ and $ \alpha^L $ corresponding to $ j \in J_K \cap J_L $ are equal, and $ a_j^K = a_j^L $, hence $ LK = 0 $, for $ j \in J_K \cap J_L $.  In this case we can rewrite Equation (\ref{2perpprod}) as:

\ba
\lefteqn{\covDff{K}{x}{\zs}} 
\AEq   \frac{ \rKnL{x-\zs}   \rKL{\aL}  - \,  \rKnL{\aK}  \rKL{\aL}  \rKnL{\zs-\zsK}    \rKL{0} }{\rKnL{x-\zs}  - \rKnL{\aK}    \rKnL{\zs-\zsK}    \rKL{0}\!^2}
\NL   \times \, \rLnK{\aL}   \covDff{K}{x^L}{\zs}    
\AEq    \rKL{\aL}   \rLnK{\aL}    \covDff{K}{x^L}{\zs}  
\AEq    \rL{\aL} \covDff{K}{x^L}{\zs}      \label{2orth_intersect}
\ea
so that
\be
\covDB{L}{K} = \rL{\aL} \covd{K}{f(\xL)}{L} \label{covKL}
\ee
We can now avoid explicit evaluation of the intractable $ \vard{K}{L}^{-1} $ term by combining Equation \eqref{covKL} with sequential update Equations \eqref{eq_BLmDK}-\eqref{eq_BLcDK} to give:
\small
\be
\ef{K \cup L}{x} = \ef{}{x} + \rK{\aK} \Df{\xK} + \rL{\aL} \Df{\xL} - \rKoL{\aLK} \Df{\xLK}  \label{AN1} 
\ee
\be
\covDff{K \cup L}{x}{x'} = \sds  \rpKoL   \R{K,L}{x}{x'}  \label{AN2}
\ee
\normalsize
\small
\be
\mbox{with} \qq \R{K,L}{x}{x'} = \sum_{i=0}^2 (-1)^i \, \sum_{T \subseteq \{K,L\}, \, |T| = i}  \re{(J_K \cup J_L) \butnot J_T}{x-x'} \re{J_T}{\aLK} \re{J_T}{\apLK} \nonumber
\ee
\normalsize
where $ J_T = \bigcup_{t \in T} J_{t} $.  Extended derivation of the general Expressions \eqref{AN1} and \eqref{AN2} for updating by any two intersecting orthogonal boundaries can be found in Appendix \ref{2PerpProof}.
Note that if $ \mk $ is defined by $ x_1 = \alpha^K $, and $ \ml $ is defined by $ x_2 = \alpha^L $, these expressions  collapse back to those given by Equations \eqref{eq_covKL1eg}.  We can see that Expressions \eqref{AN1} and \eqref{AN2} are invariant under the interchange of the two boundaries.  This should be as expected, since the boundaries are orthogonal to and intersecting each other.


\subsubsection{Two Parallel Boundaries \label{ssec_two_para_bound}}

Consider now that 
$ \ml $ is such that  $ J_K \subseteq J_L $.
In other words, $ \ml $ is either a hyperplane which is parallel to $ \mk $, or a subplane thereof.  
Note that now $ \xL = \xKL \neq \xLK $, that is, the order of the boundaries matters, and that $ \xLK \neq \xK $ in general (unless $ J_K = J_L $). We also define $ LK $ to be the $p$-vector of shortest distance from (any point on) $ \ml $ to $ \mk $.
In this case, Equation \eqref{2perpprod} can be rewritten as
\ba
\covDff{K}{x}{\zs} & = & \frac{\rK{\aL} - \rK{\aK} \rK{LK}}{1 - \rK{LK}} \rLnK{\aL}  \covDff{K}{\xL}{\zs}
\AEq \frac{\RK{\aK}{LK}}{\RK{LK}{LK}}  \rLnK{\aL}   \covDff{K}{\xL}{\zs}
\ea
hence we have that:
\be
\covd{K}{f(x)}{L }  =  \frac{ \R{J_K}{a^K}{LK} }{ \R{J_K}{LK}{LK} } \re{J_L \butnot J_K}{a^L} \covd{K}{f(x^L)}{L} \label{KxL}
\ee
Equation \eqref{KxL} allows us to again avoid explicit evaluation of the intractable 
$ \vard{K}{L}^{-1} $ term.  The adjusted expectation and covariance can then be calculated using the 
sequential update Equations \eqref{eq_BLmDK}-\eqref{eq_BLcDK}, to be:
\ba
 \ed{K \cup L}{f(x)}
  &=&  \ed{K}{f(x)} + \frac{ \R{J_K}{a^K}{LK}}{\R{J_K}{LK}{LK}} \re{J_L \butnot J_K}{a^L} (f(x^L) - \ed{K}{f(x^L)} )  \nonumber  \\
    & = & \e{f(x)} + \raK \Delta f(x^K) + \frac{ \R{J_K}{a^K}{LK}}{\R{J_K}{LK}{LK}} \re{J_L \butnot J_K}{a^L} \Delta f(x^L) \nonumber \\
    && \quad \;-\;  \frac{ \R{J_K}{a^K}{LK}}{\R{J_K}{LK}{LK}} \re{J_L \butnot J_K}{a^L} \re{J_K}{KL} \Delta f(x^{LK})  \label{Ex_two_para} \\
\covd{K \cup L}{f(x)}{f(x')} & = & \sds \, \re{P^{K \cup L}}{x-x'} R^{(2)}_{K,L}(x,x')       \label{Cov_two_para}
\ea
where we define: 
\be
R^{(2)}_{K,L}(x,x') = \RaK \re{J_L \butnot J_K}{x-x'} \;-\; \frac{ \R{J_K}{a^K}{LK} \R{J_K}{LK}{a'} }{ \R{J_K}{LK}{LK} } \re{J_L \butnot J_K}{a^L} \re{J_L \butnot J_K}{\apL} \nonumber
\ee  
Extended derivation of Expressions \eqref{Ex_two_para} and \eqref{Cov_two_para} can be found in Appendix \ref{App:2parB}.

We observe that, for the case when $ J_K \subset J_L $, the result is not invariant under the interchange of the two boundaries $ \mk \leftrightarrow \ml $, \ian{as expected}.  Although the order in which we update by the two boundaries should not affect the final result, whilst we were able to provide the analytical solution above for the case where we updated by the boundary of largest dimension first, this is not the case if we first update by the boundary of lower dimension.  As discussed in Section \ref{ssec_two_KBs}, a problem arises in the latter case due to 
us being unable to write $ \covd{K}{f(x)}{f(\zs)} $ as a product of $ \covd{K}{f(x^L)}{f(\zs)} $ and a function of $ x $ only.
Therefore, we cannot obtain an expression analogous to Equation (\ref{eq_covfxK}) which enables analytic updating of $ \simulator(x) $ by $ \mk $ and $ \ml $ by avoiding the explicit inversion of $ \vard{K}{L}^{-1} $.  In the case when $ J_K = J_L $, the result is invariant \ian{under $ \mk \leftrightarrow \ml $}, as shown in Appendix \ref{invariance_KL}.  In addition, 
if $ \mk $ is given by $ x_1 = \alpha^K $ and $ \ml $ is given by $ x_1 = \alpha^L $, Expressions \eqref{Ex_two_para} and \eqref{Cov_two_para} collapse to the result in \cite{KBECCM}.  

\subsection{Multiple Known Boundaries \label{ssub_MKBs} }


Following Sections \ref{ssec_two_orth_bound} and \ref{ssec_two_para_bound}, it is logical to assume that analytic updating would be possible for further intersecting orthogonal and parallel hyperplanes along which model behaviour is known.  
We hence proceed to discuss the form of an emulator updated by $ h $ boundaries $ \mk_H = \seqsub{\mk}{1}{h} $, where boundary $ \mk_i $ is of dimension $ p - k_i $ such that the $ x_{J_i} = x_{J_{K_i}} = \seqsub{x}{j_{i,1}}{j_{i,k_i}} $ directions are normal, with location defined by $ x_{J_i} = \alpha^{K_i} $.  In Section \ref{ssec_mult_orth_bound}, we consider intersecting orthogonal boundaries, and in Section \ref{Mult_para_bound}, we consider parallel boundaries.   Similar to the previous sections, we define \shortdist{K_i} to be the vector of shortest distance from boundary $ \mk_i $ to $ x $, where $ x^{K_i} $ is the orthogonal projection of $ x $ onto boundary $ \mk_i $.   Letting $ T = (\seqsub{t}{1}{\tau}) $ be a sequence of boundary indices, we also define $ \mk_T $ to be the sequence of boundaries $ \seqsub{\mk}{t_1}{t_\tau} $, $ \xKT $ to denote $ x $ projected sequentially onto boundaries $ \mk_T $ in reverse order (that is, $ \mk_{t_\tau}, \mk_{t_{\tau - 1}}, $ etc.), and  \shortdist{K_T}.

\subsubsection{Multiple Intersecting Orthogonal Boundaries \label{ssec_mult_orth_bound}}

In this section, we consider that the $ h $ boundaries are all orthogonal to and intersecting each other, in other words that $ \mk_1 \capdots \mk_h \neq \emptyset $.

\textbf{Theorem:} The expectation and covariance of $ \simulator(x) $ sequentially adjusted by boundaries $ \mkH $, such that $ \mk_1 \capdots \mk_h \neq \emptyset $, are given by:
\ba
\ed{\KH}{f(x)} & =& \e{f(x)} \;+\;  \sum_{i=1}^h \, (-1)^{i+1} \, \sum_{T \subseteq H, \, |T| = i} \, \rKT \Delta f(\xKT)  \label{ExKK} \\
\covd{\KH}{f(x)}{f(x')} & = & \sds \RH \, \rp{H} \label{CovKK}
\ea
\be
\mbox{with}  \quad \RH = \sum_{i=0}^h (-1)^i \sum_{T \subseteq H, \, |T| = i} \rHnT{x-x'} \rT{\akH} \rT{\apkH}  \nonumber
\ee 
where $ H = \seq{1}{h} $, 
$ J_T = \bigcup_{t \in T} J_{K_t} $, $ J_H = \bigcup_{i \in H} J_{K_i} $ and $ P^H = P \butnot J_H $.

Expressions \eqref{ExKK} and \eqref{CovKK} provide the general form for analytically updating our emulator by multiple intersecting orthogonal boundaries.
We can see that Expressions (\ref{ExKK}) and (\ref{CovKK}) are invariant under the interchange of the $ h $ boundaries.  This should be as expected, since all boundaries are orthogonal to and intersecting each other.  
Proof of Expressions (\ref{ExKK}) and (\ref{CovKK}) by induction is presented in Appendix \ref{Perp_proof}.

\subsubsection{Multiple Parallel Boundaries \label{Mult_para_bound} }

In this section, we consider that the $ h $ boundaries are such that $ J_{K_{i-1}} \subseteq J_{K_i} $.  In other words, for all $ i \geq 2 $, $ \mk_i $ is either a hyperplane which is parallel to $ K_{i-1} $, or a subplane thereof.  Such ordering of the boundaries by decreasing dimension size is required in order to leave the correlation structure in the appropriate product form to perform all the calculations analytically at each stage (see the discussion in the main part of Section \ref{ssec_two_KBs} and at the end of Section \ref{ssec_two_para_bound} for more detail).

\textbf{Theorem:} The expectation and covariance of $ \simulator(x) $ adjusted by $ h \geq 2 $ \ian{parallel} boundaries $ \mkH $, with $ J_{K_{i-1}} \subseteq J_{K_i} $ for $ i = \seq{2}{h} $,  are given by:
\small
\ba
\lefteqn{\ef{\KH}{x}}   \label{Epar}  \\ & = & \e{f(x)} + \re{J_1}{a^{K_1}} \Df{x^{K_1}}   + \sum_{\gamma=2}^h \frac{ \Rg{\gmm}{\KGm}{x}{\Kg} }{ \Rg{\gmm}{\KGm}{\Kg}{\Kg}} \rgnGm{\aKg} \bigg( \Df{\xKg} + 
 \nonumber \\
&&  \sumig \,\, \sumbG (-1)^{i+1}
   \prodli \frac{ \Rg{\blm}{\KBlm}{\Kg}{\Kbl} }{ \Rg{\blm}{\KBlm}{\Kbl}{\Kbl} } 
\rblnBlm{\Kblp \Kbl}  \Df{\xKb} \bigg) \nonumber
\ea
\ba 
\!\!\!\!\!\!\!\!\!\!\!\!\!\! \covDff{\KH}{x}{x'} & = & \sds \rp{H} \Rg{h}{\KH}{x}{x'} \label{Covpar}
\ea
\normalsize
where $ \Gamma - 1 = \seq{1}{\gamma-1} $, $ B_l = \seq{1}{b_l} $, $ B_l - 1 = \seq{1}{b_l - 1} $, $ r_\emptyset(\cdot) = 1 $, $ K_{i_1} K_{i_2} $ is the $ p $-vector of shortest distance from $ \mk_{i_1} $ to $ \mk_{i_2} $, and $ R^{(\gamma)} $ is defined recursively by:
\small
\ba
\Rg{\gamma}{\KG}{x}{x'} & = &  \bigg( \Rg{\gmm}{\KGm}{x}{x'}    \rgnGm{x-x'} 
\NL \;-\; \frac{ \Rg{\gmm}{\KGm}{x}{\Kg} \Rg{\gmm}{\KGm}{x'}{\Kg} }{ \Rg{\gmm}{\KGm}{\Kg}{\Kg} }   \rgnGm{\aKg}     \rgnGm{\apKg}    \bigg) \nonumber
\ea
\normalsize
with $ R^{(0)} = 1 $, and defining $ \Rg{\gamma}{\KG}{x}{K_i} = \Rg{\gamma}{\KG}{x}{y} $, $ \Rg{\gamma}{\KG}{K_i}{K_{i'}} = \Rg{\gamma}{\KG}{y}{y'} $ for any points $ y \in \mk_i, y' \in \mk_{i'} $ respectively.
Note that for a single boundary $ R^{(1)}_{K}(x,x') = \RaK $.
Expressions (\ref{Epar}) and (\ref{Covpar}), proved by induction in Appendix \ref{MultParB}, provide the general formulae for analytically updating our emulator by multiple parallel boundaries.  Expressions (\ref{Epar}) and (\ref{Covpar}) are not invariant under interchange of the $ h $ boundaries due to the need for the boundaries to be taken in order of decreasing dimension size in order for the calculations to be performed analytically.  


\subsection{Additional Sets of Known Boundaries \label{PerpSetsPar} }

Section \ref{ssec_mult_orth_bound} demonstrated that analytic update calculations are possible given a set of mutually orthogonal known boundaries.  Section \ref{Mult_para_bound} demonstrated that analytic update calculations are possible for sets of boundaries where each boundary is a hyperplane which is parallel to the previous one, or a subset thereof.  Given these results, the natural question to ask is: for which combinations of \ian{known} boundaries can an emulator be updated, whilst allowing all of the necessary calculations to be performed analytically?    We now state the following proposition to answer \ivm{this general question}.

\textbf{Proposition:} Given that beliefs about model output have been analytically updated given information on a sequence of boundaries $ \mk_{H-1} = \seqsub{\mk}{1}{h-1} $, with locations defined by $ \mk_i: x_{J_i} = \alpha^{K_i} $, we can update by a further boundary $ \mk_h $ if and only if, for each $ i = \seq{1}{h-1} $, either $ \mk_h \cap \mk_i \neq \emptyset $ or $ J_i \subseteq J_h $.

In other words, for each $ i = \seq{1}{h-1} $, $ \mk_h $ must either be an intersecting and orthogonal hyperplane to $ \mk_i $, or be a hyperplane (or subplane thereof) which is parallel to $ \mk_i $.

As discussed in Section \ref{ssec_two_KBs}, in order to perform an analytical update by a further known boundary, we needed to be able to write  $ \covDff{K}{x}{\zs} $ as a product of $ \covDff{K}{x^L}{\zs} $ and a function that does not depend on $ \zs $.  In other words, we needed an appropriate product correlation structure.  This same criterion extends to further known boundaries, and must hold between every pair of a set of boundaries for analytic update to be performed.  The general formula is somewhat complex, although the analytic update formulae can be derived by iteratively applying the sequential update Equations \eqref{eq_BLmDK}-\eqref{eq_BLcDK} with appropriate analogous equations to Equation \eqref{eq_covfxKpre}.

\subsection{Three-dimensional Example \label{TDE} }

\begin{figure}
\begin{center}
\includegraphics[width=13.5cm]{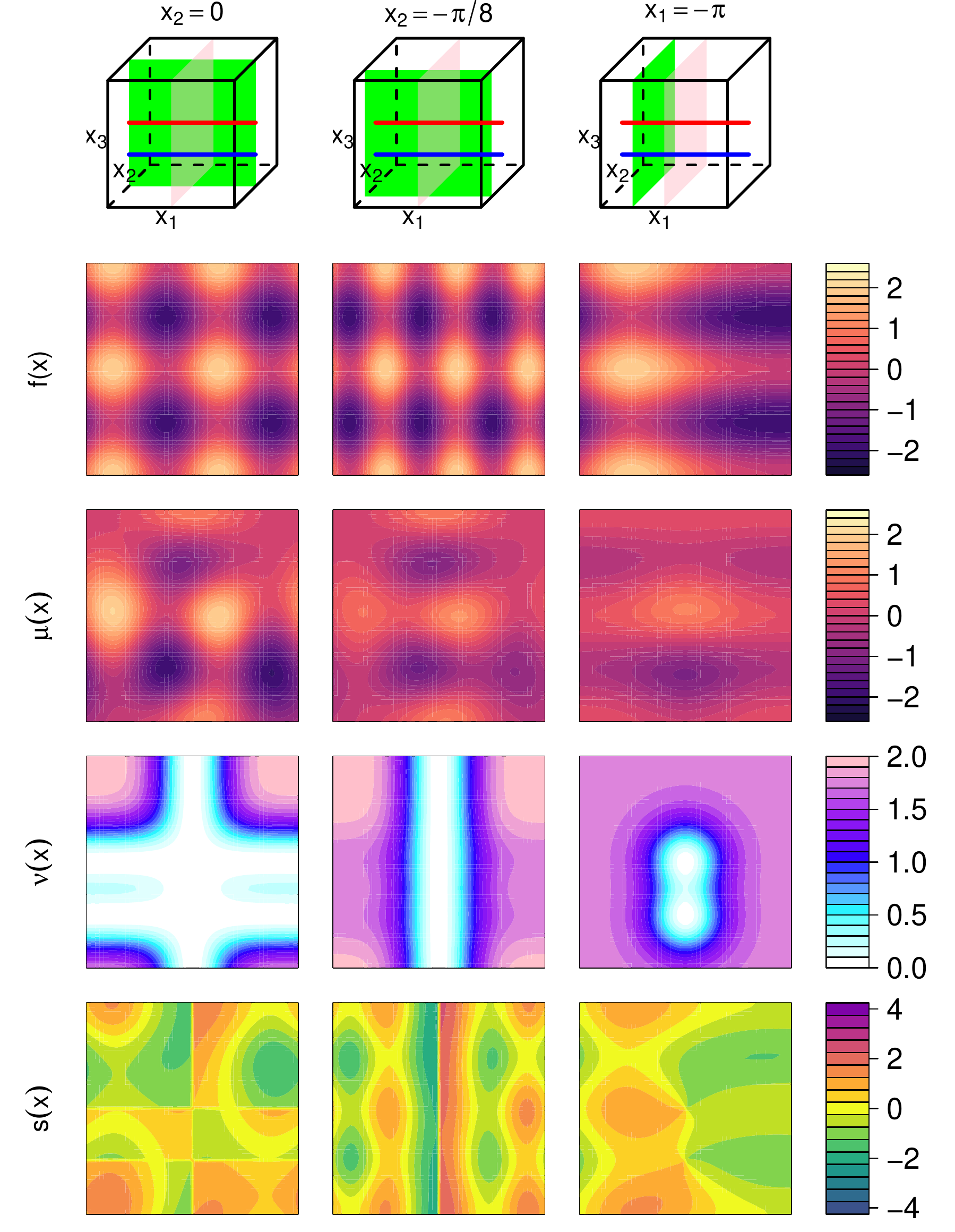}
\end{center}
\spacingset{1}
\caption{Updating the emulator for the three-dimensional function given by Equation (\ref{TEq}) by two sets of perpendicular boundaries with two and one boundaries respectively.  Rows from top to bottom, show 1) position of known boundary (red line $\mk$, blue line $\ml$ and pink plane $\mm$) and two-dimensional slices (green planes) over which the remaining plots are shown, 2) simulator function \ian{$f(x)$}, 3) emulator expectation \ian{$\mu(x)$}, 4) emulator variance \ian{$\nu(x)$}, 5) standardised errors \ian{$s(x)$}.  \ian{Columns from left to right show results on the three planes $ x_2 = 0 $, $ x_2 = -\pi/8 $ \ian{and} $ x_1 = -\pi $ respectively}. Note that for each two-dimensional plot, the variable with smaller index is along the horizontal axis. \label{3BEx} }
\end{figure}

We continue the example of Section \ref{TE} by adding two extra boundaries.  We add a \ian{one-dimensional} boundary $ \ml : (x_2,x_3) = (0, -\pi) $ which is parallel to the first, and also a \ian{two-dimensional} boundary $ \mm: x_1 = 0 $, which is orthogonal to the others. \iv{$\mk, \ml$ and $\mm$ therefore form a set of known boundaries satisfying the conditions given in the propostion in Section~\ref{PerpSetsPar}.}  The update formulae for this particular set of boundaries can be given as:
\ba
\ef{K \cup L \cup M}{x} & = & \ef{}{x} + \ro{\aM} \Df{\xM} + \rtt{\aK} \big( \Df{\xK} - \ro{\aM} \Df{x^{MK}} \big)
\NL \!\!\!\!\!\!\!\! + \, \frac{\Rtt{\aK}{LK}}{\Rtt{LK}{LK}}  \big( \Df{\xL} - \ro{\aM} \Df{x^{ML}} \big)
\NL \!\!\!\!\! - \, \frac{\Rtt{\aK}{LK}}{\Rtt{LK}{LK}} \rtt{LK} \big( \Df{\xLK} - \ro{\aM} \Df{x^{MLK}}  \big) \label{EgE}  \\
\vard{K \cup L \cup M}{f(x)} & = & \sds \Rg{2}{K,L}{x}{x}   \R{1}{\aM}{\aM}  \label{EgV}
\ea
The derivation of Equations \eqref{EgE} and \eqref{EgV} is provided in Appendix \ref{Eg_App}.
The emulator outputs, derived using these equations, are shown in Figure~\ref{3BEx}.
We can see that with these three boundaries, much is learnt across each of the displayed two-dimensional slices of the input space.  Variance is particularly reduced for $ x_2 = 0 $ (left-hand column), 
this column also essentially containing the story of a smaller two-dimensional example (that when $ x_2 = 0 $) with three one-dimensional boundaries.  The emulator predicts the model across much of the input space well; 
only in the top left and top right corners, when $ x_3 $ is large and $ x_1 $ small or large, is behaviour really uncertain.

The middle column shows the plane $ x_2 = \pi/8 $.  We can see that the intersecting known boundary \ian{$ \mm$ at }$x_1 = 0 $ has much greater influence on the adjusted beliefs across the plane of interest than the lower dimensional known boundaries \ian{$\mk$ and $\ml$}, these being subplanes of a plane parallel to the one of interest in this case.  In contrast, if the plane of interest is parallel to the two-dimensional plane, for example $ x_1 = \pi $ in the right-hand column, then the intersecting lines have a greater influence, although concentrated over a smaller area of the plane.  The right-hand column particularly highlights the advantages of having as many known boundaries as possible.  The intersecting lines provide much increased precision over a smaller area, whilst the parallel plane reduces variance slightly (though still to a worthwhile degree) across the whole plane.  In addition, the diagnostics are satisfactory across each plane in the example.

To summarise, for computer model applications where such sets of known boundaries exist, the gains of including them in the analysis using the general results derived in this section can be substantial, \iv{and therefore they should be included whenever possible}.


\section{Application of Methods to Arabidopsis Model \label{AMAM} }

In the previous sections of this article, we have presented methodology for utilising knowledge of the behaviour of computer models along particular boundaries of the input space to aid emulation across the whole input space.  In this section, we explore the implications such boundaries can have on a higher-dimensional scientifically relevant systems biology model of the hormonal crosstalk in the roots of an Arabidopsis plant.

\subsection{Model of Hormonal Crosstalk in Arabidopsis Thaliana}

\emph{Arabidopsis Thaliana} is a small flowering plant that is widely used as a model organism in plant biology \citep{AAT}. 
We demonstrate our known boundary emulation techniques on a model of hormonal crosstalk in the root of an Arabidopsis plant that was constructed by \cite{IPPHCARD}.  This Arabidopsis model represents the crosstalk of auxin, ethylene and cytokinin in Arabidopsis root development as a set of 18 differential equations, given in Table \ref{DE} of Appendix \ref{App:AMS}, which must be solved numerically.  The model takes an input vector of 45 rate parameters $ (k_1, k_{1a}, k_2, ...) $, although we will be interested in a subset of 38 of them, as discussed in Appendix \ref{App:AMS}, and returns an output vector of 18 chemical concentrations $ ([Auxin], [X], [PLSp], ... ) $.  This Arabidopsis model has been successfully emulated in the literature in the context of history matching \citep{BUCSBM}. 

For the purposes of this article, we are interested in modelling the important output component $ [ET] $, which represents the concentration of ethylene \citep{EPG, EUAB}, at early time $ t = 2 $.  The ranges over which we allowed the inputs to vary are given in Table \ref{RR} in Appendix \ref{App:AMS}, these being elicited as ranges of interest deemed sensible by the biological experts \citep{IPPHCARD}, and square rooted and mapped to a $ [-1,1 ] $ scale prior to analysis.

\subsection{Establishing Known Boundaries \label{EKB} }

Establishing known boundaries requires some understanding of the scientific model. 
It is not uncommon for one or more known boundaries to occur in a model for some output components. Often, setting certain parameters to
specific values will decouple smaller subsections of the system, which may allow subsets of the model equations to be solved analytically, for particular output components, as is the case for the Arabidopsis model.

We consider known boundaries for output component $ [ET] $ by considering its rate equation:
\begin{equation}
\frac{d[ET]}{dt} \;=\; k_{12} + k_{12a}[Auxin][CK] - k_{13}[ET]
\end{equation}
A known boundary exists when rate parameter $ k_{12a} = 0 $, since in this case:
\begin{equation}
\frac{d[ET]}{dt} \;=\; k_{12} - k_{13}[ET] \;\;\;\; \Rightarrow \;\;\;\; [ET] \;=\; \frac{([ET^0]k_{13} - k_{12})  \exp(-k_{13}t) + k_{12}}{k_{13}}
\end{equation}
where $ [ET^0] $ is the initial condition of the $ [ET] $ output component, and we see that $ [ET] $ has been entirely decoupled from the rest of the system.  $ [ET] $ can now be obtained along the boundary $ k_{12a} = 0 $ with negligible computational cost.  Note that this boundary is of dimension $ p-1 = 38 -1 = 37 $.  The second (perpendicular) known boundary for $ [ET] $ is a $ p-4 = 34 $-dimensional boundary given by $ k_{1a}  = k_{2a} = k_{3a} = k_{18a} = 0 $, which \ian{decouples} the combined system of $ [Auxin], [CK] $ and $ [ET] $.  In this case, we can solve for $ [Auxin] $ and $ [CK] $ first:
\begin{eqnarray}
\frac{d[Auxin]}{dt} \;&=&\; k_{2} - k_{3}[Auxin]  \nonumber \\
\Rightarrow \;\;\;\; [Auxin] \;&=&\; \frac{([Auxin^0]k_{3} - k_{2})  \exp(-k_{3}t) + k_{2}}{k_{3}} \\
\frac{d[CK]}{dt} \;&=&\; - k_{19}[CK]  \nonumber  \\
\Rightarrow \;\;\;\; [CK] \;&=&\; [CK^0] \exp(-k_{19}t) 
\end{eqnarray}

Inserting these solutions into the rate equation for $ [ET] $ then yields:
\begin{eqnarray}
\frac{d[ET]}{dt} \;&=&\; k_{12} + k_{12a}[CK^0] \exp(-k_{19}t) \left( \frac{([Auxin^0]k_3 - k_2) \exp(-k_3t) + k_2}{k_3} \right) - k_{13}[ET]  \nonumber \\
\Rightarrow \; [ET] \;&=&\; \frac{k_{12}}{k_{13}} (1 - \exp(-k_{13}t) ) + \frac{k_{12a}[CK^0]k_2}{k_3(k_{19}-k_{13})} \big(1 - \exp \big((k_{13} - k_{19})t\big)\big) \nonumber \\
&& \quad  \;+\;  \frac{k_{12a}[CK^0]([Auxin^0]k_3 - k_2)}{k_3(k_3 + k_{19} - k_{13})} \big(1 - \exp \big((k_{13} - (k_3 + k_{19}))t\big) \big) 
\end{eqnarray}
which can now be solved analytically with negligible computational cost, given the initial conditions $[Auxin^0]$ and $[CK^0]$ for Auxin and Cytokinin respectively.  In this case, we have $ [CK^0] = [ET^0] = [Auxin^0] = 0.1 $ as the initial conditions suggested by the biological experts.  The remaining initial conditions are shown in Table \ref{IOV} in Appendix \ref{App:AMS}.  We will refer to this $ p-4 $-dimensional boundary as $ \mk $ and the earlier presented $ p-1 $-dimensional boundary as $ \ml $ in order to show the effect of the smaller-dimension boundary in comparison to the larger-dimension one.  In addition, it is important to note that both boundaries $ \mk $ and $ \ml $ lie outside the input space of interest $ \mx $ as given by Table \ref{RR} in Appendix \ref{App:AMS}.  Despite this, assuming the behaviour of the model is reasonable \ian{in the vicinity of} the boundaries, the information provided by the analytical solutions along the boundary can be useful for predicting model behaviour inside $ \mx $.

\subsection{Emulator Structure and Specification \label{ESS}}

We restrict the form of our emulator to have the covariance structure as given by Equation \eqref{eq_cor_struc}.  We used a product Gaussian correlation function of the form given by Equation (\ref{GF}), as we assumed that the solution to the Arabidopsis model would most likely be smooth and that many orders of derivatives would exist.

The prior emulator expecation and variance were taken to be constant, that is $ \e{f(x)} = \beta $ and $ \var{f(x)} = \sds $, where $ \beta $ and $ \sds $ were estimated to be the sample mean and variance of a set of previously evaluated scoping runs.  In this section, we specify a common correlation length parameter $ \theta = 3 $ for each input, a choice consistent with the argument for approximately assessing correlation lengths presented in \cite{GFBUA}.  This value of $ \theta $ was also checked for adequacy using standard emulator diagnostics \citep{DGPE}.  We made this relatively simple emulator specification for illustrative purposes, the reason being that we wish to demonstrate that there are benefits to utilising the known boundaries regardless of how the parameters \ian{may have been estimated}.  To this end, in Section \ref{SEPS} we compare the effects of several different values of $ \theta $ on an analysis with and without the known boundaries, but for now keep the value fixed at $ \theta = 3 $.

\subsection{Comparison of Results \label{CR}}

In this section, we compare the emulators of the above form constructed with and without use of the known boundaries $ \mk: k_{1a}  = k_{2a} = k_{3a} = k_{18a} = 0 $ and $ \ml: k_{12a} = 0 $, and also with and without the addition of training points.  The design for the additional training points is obtained by constructing a Maximin Latin hypercube design of size 1000 across the 38-dimensional input space, this then being sampled from to explore the effects of using different numbers of training points up to 1000.  Bayes linear updates were carried out using the single and two perpendicular boundary updates given by Equations (\ref{eq_EK1}), (\ref{eq_covK2}) and (\ref{ExKK}), (\ref{CovKK}) respectively.  Additional updating is then performed using the sequential update formulae given by Equations (\ref{eq_BLmDK})-(\ref{eq_BLcDK}).

\begin{figure}
\begin{center}
\includegraphics[width=15.7cm]{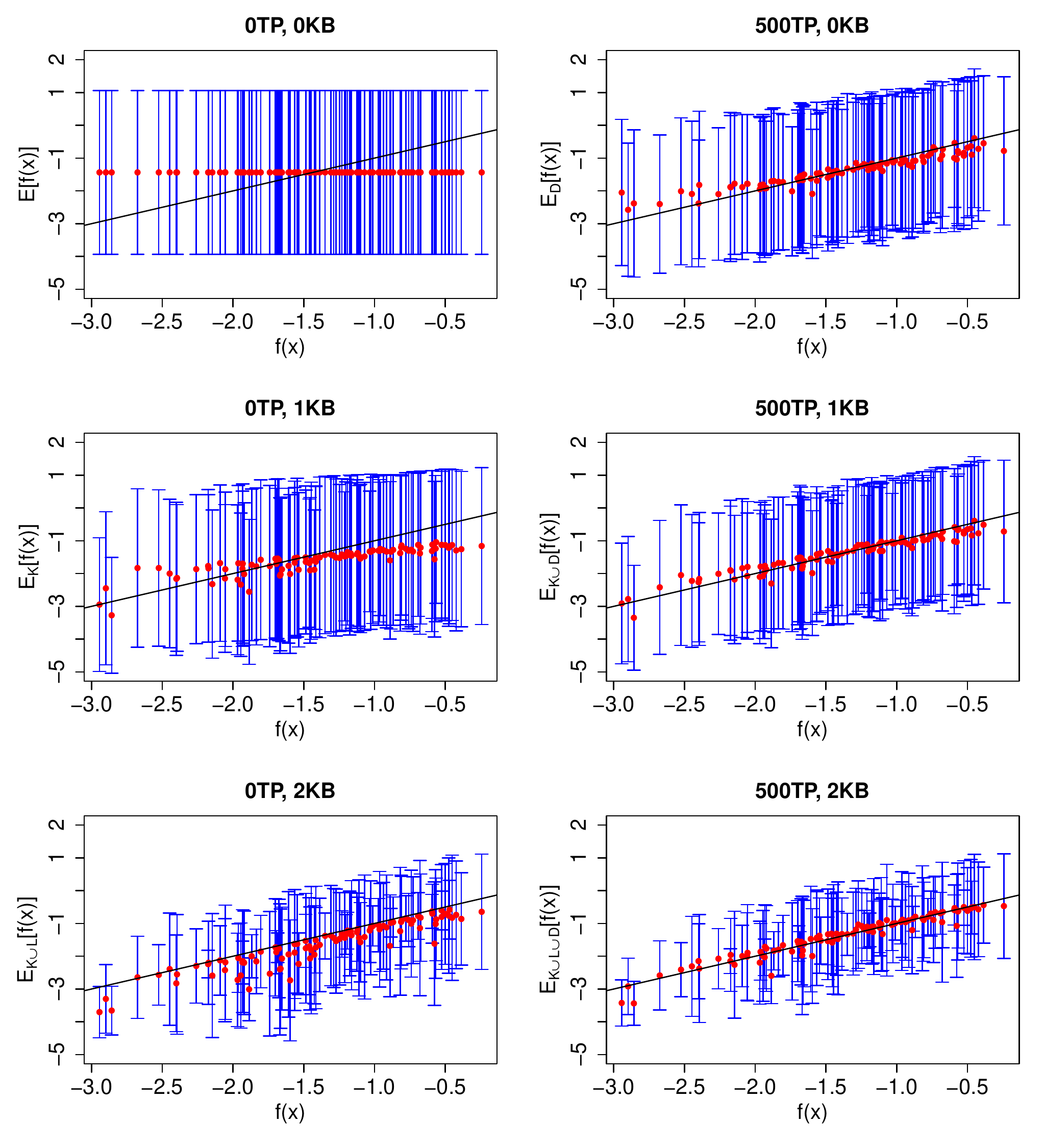} 

\spacingset{1}
\caption[Updating by Known Boundaries for Arabidopsis]{Diagnostic plots for the emulators of the Arabidopsis model output component $ [ET] $.  These show true model output against emulator expectation plus/minus 3 standard deviations for a set of diagnostic test points given; first row - no known boundary; second row - single known boundary $ \mk $; third row - two known boundaries $ \mk $ and $ \ml $.  The left column shows diagnostics for emulators without additional training points $ D $ in the bulk of the input space, and the right column shows with 500 additional points, all for common correlation length parameter $ \theta = 3 $ for all input parameters.}
\label{ArabEg}
\end{center}
\end{figure}

Equivalent plots to those shown in Figures \ref{1BEx} and \ref{3BEx} are substantially more difficult to visualise across all dimensions of a high-dimensional input space.  We will use numerical diagnostics to assess these emulators in Section \ref{SEPS}, but in this section we will restrict comparison of the emulators to visual diagnostics.  Figure \ref{ArabEg} shows model output against emulator expectation $ \pm 3 $ standard deviations for a set of 100 diagnostic test points for each of six emulators; first row: no boundaries; second row: 1 known boundary $ \mk $; third row: two known boundaries $ \mk $ and $ \ml $.  The left column shows diagnostics for emulators without additional training points $ D $ in the bulk of the input space, and the right column shows with 500 additional points.
If the error bars for too many points do not intersect the line $ f(x) = \e{f(x)} $, this would suggest that the emulator is not valid.  This heuristic appeals to Pukelsheim's Three Sigma Rule \citep{3SR} which states that 95\% of the probability mass of any unimodal distribution lies within 3 standard deviations of the mean.

The middle left panel of Figure \ref{ArabEg} shows 
that the expected values of the points have been marginally influenced in general by $ \mk $, but to a greater degree for inputs for which the model output is smaller.  The bottom left panel shows 
that the addition of $ \ml $ results in a much \iv{improved} effect on the predictions than boundary $ \mk $ alone, 
We also notice that utilising
the known boundaries only (without training points) results in (slightly) underestimated predictions, however, the diagnostics are still satisfactory.  In addition, the results are comparable to using no boundaries and 500 training points across the input space (top right panel), thus highlighting that utilising knowledge of computer model behaviour along known boundaries is worthwhile.   Crucially, however, whereas the 500 training points require 500 potentially computationally intensive model evaluations and emulator matrix inversion calculations, the known boundaries involve no model evaluations or matrix inversion calculations.  

The substantial and moderate effects of $ \ml $ and $ \mk $ respectively on our beliefs 
in comparison to individual points is largely a result of the dimension of the objects.  The known boundaries are \iv{$ p-1 $- and $ p-4 $}-dimensional objects respectively, resulting in significant variance resolution as a consequence of the volume of the input space within their proximity (particularly $ \ml $ for which the correlation function is effectively over a single dimension). In comparison, individual training runs (which are $0-$dimension\iv{al} objects) \iv{influence far smaller volumes} especially in high dimensions.  

Since there is little computational cost involved in the incorporation of known boundaries, the most practical solution is to utilise them in conjunction with the regular training points.   Looking at the bottom right panel of Figure \ref{ArabEg}, 
we notice a substantial improvement in comparison to using either the known boundaries or the training points individually. \ian{Were one aware of the known boundaries in advance, one could design the set of 500 runs accordingly, leading to further efficiency gains (see~\cite{KBECCM}).}

\subsection{Sensitivity to Emulator Parameter Specification \label{SEPS}}

We now compare emulators constructed using various different emulator parameter specifications.  In particular, we explore the effect of changing the common correlation length parameter $ \theta $ discussed in Section \ref{ESS}.  We do this as we wish to focus on the advantage of utilising known boundaries on emulation without confounding the effect on choice of parameter specification.  Whilst we will demonstrate that the effects of known boundaries are substantial regardless of emulator structure and parameter specification, the value of $ \theta $ does \ian{affect the relative size of the contributions} of individual points to known boundaries (that is, larger dimensional objects).   We compare several emulators with various values of correlation length parameter $ \theta $, numbers of training points and numbers of known boundaries using numerical diagnostics for 500 diagnostic points.  
These diagnostics, shown in Table~\ref{tab:SumVarMASPE},
are the sum of variances and Mean Absolute Standardised Prediction Error (MASPE), given by:
\begin{equation}
\sum_{w=1}^{500} \nu(x^{(w)})   \qq \mbox{and} \qq \frac{1}{500} \sum_{w=1}^{500} \frac{|f(x^{(w)}) - \mu(x^{(w)})|}{\sqrt{\nu(x^{(w)})}} \nonumber
\end{equation}
respectively, where $ \mu(x) $ and $ \nu(x) $ represent the appropriate emulator mean and variance in each case.

The prior sum of variances is 344.23 (constant for all $ \theta $), this being reduced by various degrees depending on the three varying features of our analysis.  For small $ \theta = 0.1 $, neither training points nor known boundaries reduce the variances of the diagnostic points appreciably.  With $ \theta = 1 $, training points are having negligible effect on variance, however, the larger known boundary objects have sufficient diagnostic points within their proximity  to reduce uncertainty to some degree.  For $ \theta = 6 $, the reduction in diagnostic variance arising from two known boundaries only (15.04) is greater than that of 1000 training points alone (23.87).  
As $ \theta $ gets larger 
1000 training points in $ \mathcal{X}  $ have greater affect than two known boundaries outside of $ \mathcal{X} $, for example, reducing the sum of variances to 2.07 and 2.44 respectively when $ \theta = 10 $ (and to 0.02 when both are used).  
\ian{These results are as expected from purely geometrical considerations}.

\begin{table}
\spacingset{1}
\footnotesize
\centering
\begin{tabular}{|c|ccccc|ccccc|}
  \hline
  & \multicolumn{5}{|c|}{Sum of Variances} & \multicolumn{5}{c|}{MASPE} \\ \hline
$ \theta $ & 0.1 & 1 & 3 & 6 & 10 & 0.1 & 1 & 3 & 6 & 10  \\ 
  \hline
0TP, 0KB & 344.23 & 344.23 & 344.23 & 344.23 & 344.23 & 0.84 & 0.83 & 0.83 & 0.83 & 0.83 \\ 
  0TP, 1KB & 344.23 & 344.08 & 284.03 & 131.07 & 55.56 & 0.84 & 0.83 & 0.60 & 1.50 & 4.26 \\ 
  0TP, 2KB & 344.23 & 295.29 & 100.77 & 15.04 & 2.44 & 0.84 & 0.67 & 1.77 & 24.67 & 182.44 \\ 
  200TP, 0KB & 344.23 & 344.23 & 277.21 & 55.00 & 9.19 & 0.84 & 0.83 & 0.52 & 1.26 & 7.35  \\ 
  200TP, 1KB & 344.23 & 344.08 & 230.32 & 24.13 & 2.12  & 0.84 & 0.83 & 0.43 & 3.61 & 49.21 \\ 
  200TP, 2KB & 344.23 & 295.29 & 81.11 & 2.71 & 0.09 & 0.84 & 0.67 & 1.00 & 60.25 & 2508.44 \\ 
  500TP, 0KB & 344.23 & 344.23 & 251.15 & 37.13 & 5.04 & 0.84 & 0.83 & 0.37 & 1.59 & 12.50 \\ 
  500TP, 1KB & 344.23 & 344.08 & 209.05 & 16.29 & 1.10 & 0.84 & 0.83 & 0.35 & 4.66 & 82.04 \\ 
  500TP, 2KB & 344.23 & 295.29 & 73.44 & 1.82 & 0.05 & 0.84 & 0.67 & 1.02 & 83.80 & 4305.17 \\ 
  1000TP, 0KB & 344.23 & 344.23 & 229.70 & 23.87 & 2.07  & 0.84 & 0.83 & 0.33 & 2.16 & 25.09 \\ 
  1000TP, 1KB & 344.23 & 344.08 & 191.42 & 10.87 & 0.56 & 0.84 & 0.83 & 0.33 & 5.86 & 129.16 \\ 
  1000TP, 2KB & 344.23 & 295.29 & 67.08 & 1.21 & 0.02 & 0.84 & 0.67 & 1.07 & 117.86 & 7764.93 \\
   \hline
\end{tabular}
\caption{\small{Sum of Variances and Mean Absolute Standardised Prediction Error for the set of 500 diagnostic points for different values of common $ \theta $ and numbers of known boundaries (KB) and training points (TP) in the bulk of the input space.} \label{tab:SumVarMASPE}}
\end{table}

It is common for acceptable values of the MASPE to be broadly around 1 (appealing to the properties of a standard half-normal distribution, which has expectation $ \sqrt{2/\pi} $), and providing substantial evidence that an emulator is invalid if much greater than 2 or 3 (appealing to Pukelsheim's $ 3\sigma $ rule \citep{3SR}).  Equivalently, substantial change in MASPE between prior and adjusted beliefs is also cause for concern.  Prior MASPE is 0.84, which is suitably close to both 1 and $ \sqrt{2/ \pi} \approx 0.8 $.  The MASPE values for emulators with large values of $ \theta $ are \iv{as expected} unacceptable, with the value for 1000 training points alone being 25.09 and that for 1000 training points and two known boundaries 7764.93.
This excessively larger value 
is due to the different ways in which known boundaries and training points influence the emulator.  The known boundaries affect the input space as a large object with much influence \iv{over} a particular part of the input space.  On the other hand, since the training points are spread out across $ \mathcal{X} $, the effect of averaging via interpolation of the points is likely to result in more accurate (and thus with common variance reduction appear more valid) predictions, even if $ \theta $ is large.  The MASPE values for $ \theta = 3 $, 1000 training points and two known boundaries is 1.07, which is much more acceptable. \iv{For the emulators with acceptable diagnostics, we see that the inclusion of known boundaries is clearly beneficial.}
In addition to sum of variances and MASPEs, we also calculated Root Mean Square Errors (RMSEs) for each emulator, these being displayed and discussed in Appendix \ref{App:AMS}.


\section{Conclusion \label{conc} }

We have discussed how additional prior insight into the physical structure of a computer model \iv{related to known boundaries} can be incorporated into emulators \iv{leading to} substantial increases in accuracy \iv{for} little additional computational cost.  

In particular, here it is shown that if a computer model has boundaries or hyperplanes in its input space where it can either be analytically solved or just evaluated far more efficiently, then these known boundaries can be formally incorporated into the emulation process by \ian{analytic} Bayesian updating of the emulators with respect to the information contained on the boundaries.  Furthermore, we have shown that this is possible for a large class of emulators, and for multiple boundaries of various forms.  The progress in this work in comparison to \cite{KBECCM} is that we presented \ian{substantially more general} results \iv{for arbitrary numbers of boundaries of varying dimension}, stating which \iv{configurations} of known boundaries permit analytical updates. \iv{Due to these analytic results and to the ease and substantial benefits of including known boundaries when emulating the Arabidopsis model, we would suggest that future UQ analyses of serious scientific models include a phase of identification and incorporation of known boundaries, if they are found to exist, as standard practice}.  Whilst the results of this article were with respect to a univariate computer model, the results extend naturally to the multivariate case, as discussed in Appendix~\ref{ME}.  

There are many ways in which the work of this article could be developed.  For example, extensions to the case of uncertain regression 
parameters \iv{within the emulator} are possible, although the formal update would now depend on the specific form of 
the correlation function $r_j(a)$, which may not be tractable for \iv{some} choices. Curved boundaries of various geometries could also be incorporated, provided both that suitable transformations were found to convert them to hyperplanes and that we were happy to adopt the induced transformed product correlation structure as our prior beliefs.
Finally we note that, for some applications, there may be several hyperplanes in the input space along which model behaviour is known, however, analytical updates incorporating the information given by all of them may not be possible due to the set not satisfying the properties of the proposition in Section \ref{PerpSetsPar}.  There is then a possible design problem which involves selecting the best (in some sense) subset  of the known boundaries which do permit analytic updating.  This choice of boundaries may be in conjunction with design of the training points in the bulk of the input space $ X_D $.  Training point design should anyway take the known boundaries into account, as discussed in \cite{KBECCM}.
We leave all these considerations to future research.


\bibliographystyle{plain}


\appendix


\section{Extended Derivation of Expressions \eqref{AN1} and \eqref{AN2} \label{2PerpProof}}

Here we provide a more detailed derivation of Expressions \eqref{AN1} and \eqref{AN2}, presented in Section \ref{ssec_two_orth_bound}.  Firstly, the expectation adjusted by $ \mk $ and $ \ml $ can be calculated, using the 
sequential update Equation (\ref{eq_BLmDK}), to be:
\small
\ba
\lefteqn{\ef{K \cup L}{x}}
\AEq \ef{K}{x} + \covd{K}{f(x)}{L}   \vard{K}{L}^{-1}   (L - \ed{K}{L})
\AEq \ef{K}{x} + \rL{\aL} \big( f(\xL) - \ef{K}{x^L} \big)
\AEq \ef{}{x} + \rK{\aK} \Df{\xK} + \rL{\aL} \bigg( f(\xL) - \bigg( \ef{}{\xL} + \rK{\xL - \xLK} \Df{\xLK} \bigg) \bigg)
\AEq \ef{}{x} + \rK{\aK} \Df{\xK} + \rL{\aL} \Df{\xL} - \rL{\aL} \rKnL{\aK} \rKL{0} \Df{\xLK}
\AEq \ef{}{x} + \rK{\aK} \Df{\xK} + \rL{\aL} \Df{\xL} - \rKoL{\aLK} \Df{\xLK} 
\ea
\normalsize
Similarly, by combining Equation \eqref{covKL} with Equation \eqref{eq_BLcDK}, we have that:
\small
\ba
\lefteqn{\covDff{K \cup L}{x}{x'}}
\AEq \covDff{K}{x}{x'}   -  \covd{K}{f(x)}{L} \vard{K}{L}^{-1}  \covd{K}{L}{f(x)} 
\AEq \covDff{K}{x}{x'} - \rL{\aL} \covDff{K}{\xL}{\xpL}  \rL{\aL}
\AEq  \sds  \RaK  \rp{K}  - \, \rara{L}  \sds \RK{\xL-\xLK}{\xpL - \xpLK}  \re{P^K}{\xL-\xpL}
%
%
\AEq   \sds \RaK  \rLnK{x-x'} \rp{K \cap L} 
\NL  - \, \rara{L}  \sds \RKnL{\aK}{\apK}  \rLnK{\xL-\xpL}  \rpKoL
\AEq   \sds \rpKoL  \bigg(  \bigg(  \rK{x-x'} - \rara{K} \bigg)  \rLnK{x-x'} 
\NL  - \,  \rara{L} \bigg( \rKnL{x-x'} - \rKnL{\aK} \rKnL{\apK} \bigg)  \bigg)
\AEq   \sds   \rpKoL   \bigg(  \rK{x-x'}  \rLnK{x-x'} - \rara{K} \rLnK{x-x'}  
\NL - \, \rara{L} \rKnL{x-x'} + \rara{L}  \rKnL{\aK} \rKnL{\apK} \bigg) 
\AEq   \sds    \rpKoL   \bigg(  \rKoL{x-x'} - \, \rK{\aLK}  \rK{\apLK} \rLnK{x-x'} 
\NL - \, \rL{\aLK}  \rL{\apLK}  \rKnL{x-x'} + \rKoL{\aLK}   \rKoL{\apLK}   \bigg)
\AEq  \sds  \rpKoL   \R{K,L}{x}{x'} 
\ea
\normalsize
where
\small
\be
 \R{K,L}{x}{x'} = \sum_{i=0}^2 (-1)^i \, \sum_{T \subseteq \{K,L\}, \, |T| = i}  \re{(J_K \cup J_L) \butnot J_T}{x-x'} \re{J_T}{\aLK} \re{J_T}{\apLK}
\ee
\normalsize


\section{Extended Derivation of Expressions \eqref{Ex_two_para} and \eqref{Cov_two_para} \label{App:2parB}}

Here we provide a more detailed derivation of Expressions \eqref{Ex_two_para} and \eqref{Cov_two_para}, presented in Section \ref{ssec_two_para_bound}.  
Firstly, the expectation adjusted by $ \mk $ and $ \ml $ can be calculated, using the 
sequential update Equation (\ref{eq_BLmDK}), to be:
\ba
 \lefteqn{\ed{K \cup L}{f(x)}} \nonumber \\
  &=& \ed{K}{f(x)} + \covd{K}{f(x)}{L} \vard{K}{L}^{-1}(L- \ed{K}{L})  \nonumber \\
 &=&  \ed{K}{f(x)} + \frac{ \R{J_K}{a^K}{LK}}{\R{J_K}{LK}{LK}} \re{J_L \butnot J_K}{a^L} \covd{K}{f(x^L)}{L} \vard{K}{L}^{-1}(L- \ed{K}{L})  \nonumber \\
  &=&  \ed{K}{f(x)} + \frac{ \R{J_K}{a^K}{LK}}{\R{J_K}{LK}{LK}} \re{J_L \butnot J_K}{a^L} (f(x^L) - \ed{K}{f(x^L)} )  \nonumber  \\
    &=&   \e{f(x)} + \raK \Delta f(x^K)   \nonumber  \\ 
    && \quad + \, \frac{ \R{J_K}{a^K}{LK}}{\R{J_K}{LK}{LK}} \re{J_L \butnot J_K}{a^L} \left\{ f(x^L) - \left(\e{f(x^L)} + \re{J_K}{KL} \Delta f(x^{LK})  \right) \right\}  \label{Ex_two_para_half} \nonumber \\
    & = & \e{f(x)} + \raK \Delta f(x^K) + \frac{ \R{J_K}{a^K}{LK}}{\R{J_K}{LK}{LK}} \re{J_L \butnot J_K}{a^L} \Delta f(x^L) \nonumber \\
    && \quad \;-\;  \frac{ \R{J_K}{a^K}{LK}}{\R{J_K}{LK}{LK}} \re{J_L \butnot J_K}{a^L} \re{J_K}{KL} \Delta f(x^{LK})  
\ea
Similarly, using Equation \eqref{eq_BLcDK}, we find the covariance adjusted by $ \mk $ and $ \ml $ to be:
\ba
\lefteqn{ \covd{K \cup L}{f(x)}{f(x')} } \nonumber \\
     &=& \covd{K}{f(x)}{f(x')} - \covd{K}{f(x)}{L} \vard{K}{L}^{-1}\covd{K}{L}{f(x')}  \nonumber \\ 
     &=& \covd{K}{f(x)}{f(x')} - \frac{ \R{J_K}{a^K}{LK}}{\R{J_K}{LK}{LK}} \re{J_L \butnot J_K}{a^L} \covd{K}{f(x^L)}{f(x')}  \nonumber \\
     &=&  \covd{K}{f(x)}{f(x')} \nonumber \\
     &&  \quad \;-\; \frac{ \R{J_K}{a^K}{LK}}{\R{J_K}{LK}{LK}} \re{J_L \butnot J_K}{a^L} \covd{K}{f(x^L)}{f(x'^L)} \re{J_L \butnot J_K}{\apL} \frac{ \R{J_K}{LK}{a'}}{\R{J_K}{LK}{LK}} \label{Cov_two_para_half} \nonumber \\
     &=& \sds \RaK \, \re{{P^K}}{x-x'}  \nonumber \\
     && \quad \;-\; \frac{ \R{J_K}{a^K}{LK}}{\R{J_K}{LK}{LK}} \re{J_L \butnot J_K}{a^L} \sds \R{J_K}{LK}{LK}  \, \re{P^{K \cup L}}{x-x'} \re{J_L \butnot J_K}{\apL} \frac{ \R{J_K}{LK}{a'}}{\R{J_K}{LK}{LK}}  \nonumber \\
    &=&  \sds \, \re{P^{K \cup L}}{x-x'}  \nonumber \\
     &&  \quad \tim\left\{ \RaK \re{J_L \butnot J_K}{x-x'} - \frac{\R{J_K}{a^K}{LK} \R{J_K}{LK}{a'}}{\R{J_K}{LK}{LK}} \re{J_L \butnot J_K}{a^L} \re{J_L \butnot J_K}{\apL} \right\} \nonumber \\ 
    &=&  \sds \, \re{P^{K \cup L}}{x-x'} R^{(2)}_{K,L}(x,x')    
\ea
where we define: 
\be
R^{(2)}_{K,L}(x,x') = \RaK \re{J_L \butnot J_K}{x-x'} \;-\; \frac{ \R{J_K}{a^K}{LK} \R{J_K}{LK}{a'} }{ \R{J_K}{LK}{LK} } \re{J_L \butnot J_K}{a^L} \re{J_L \butnot J_K}{\apL} \nonumber
\ee


\section{Invariance for Two Parallel Boundaries of Equal Dimension \label{invariance_KL}}

We here demonstrate the invariance of the two parallel boundary case when the boundaries are of equal dimension, that is $ J_K = J_L $.  In this case, Expression (\ref{Ex_two_para}) reduces to:
\small{
\ba
\ed{K \cup L}{f(x)} &=& \e{f(x)} + \raK \Delta f(x^K)  \nonumber \\
   && \qq +  \frac{ \R{J_K}{a^K}{LK}}{\R{J_K}{LK}{LK}} \left\{ f(x^L) - \left(\e{f(x^L)} + \re{J_K}{LK} \Delta f(x^K)  \right) \right\} 
\ea}
\normalsize
which can then be written as:
\ba
 \lefteqn{\ed{K \cup L}{f(x)} } \nonumber \\
 &=&  \e{f(x)} + \left( \raK - \frac{ \R{J_K}{a^K}{LK}}{\R{J_K}{LK}{LK}} \re{J_K}{LK} \right) \Delta f(x^K)    + \, \frac{ \R{J_K}{a^K}{LK}}{\R{J_K}{LK}{LK}} \Delta f(x^L) \nonumber  \\  
   &=& \e{f(x)} + \left[ \frac{\raK- \re{J_L}{\aL}\re{J_K \cap J_L}{LK}}{1-\re{J_K \cap J_L}{LK}\!^2 } \right] \Delta f(x^K)  \nonumber \\
   && \qq \qq \qq + 
		\left[\frac{\re{J_L}{\aL} - \raK\re{J_K \cap J_L}{LK}}{1-\re{J_K \cap J_L}{LK}\!^2 }\right] \Delta f(x^L) \nonumber \\
		\label{eq_expLK1_para}
\ea
where we have exploited the fact that the projection of $ x^L $ onto $ \mk $ is just $ x^K $, and that $ J_K = J_L = J_K \cap J_L $.  Expression (\ref{eq_expLK1_para}) is explicitly invariant under the interchange of the two boundaries $ \mk \leftrightarrow \ml $ (as $LK=\aK+\aL$ is invariant under $\aK \leftrightarrow \aL, \apK \leftrightarrow \apL $).

Similarly, Expression \eqref{Cov_two_para} reduces to:
\be
\covd{K \cup L}{f(x)}{f(x')} = \sds \, \re{{P^K}}{x-x'} \bigg( \RaK - \frac{\R{J_K}{a^K}{LK} \R{J_K}{LK}{a'}}{\R{J_K}{LK}{LK}}  \bigg)
\ee
Therefore we obtain:
\ba
\lefteqn{\covd{K \cup L}{f(x)}{f(x')}} \nonumber \\
  &=&  \sds \, \frac{\re{{P^K}}{x-x'}}{\R{J_K}{LK}{LK}} \bigg( 
    	\big( \raaK- \raK  \rapK \big) \big(  1 -   \re{J_K}{LK}\!^2  \big) \;   \nonumber \\
    &&   \quad\quad \quad \quad \quad \quad \quad \quad  \;-\; \big(\re{J_K}{\aL} -  \raK \re{J_K}{LK}  \big)\big( \re{J_K}{\apL} -   \re{J_K}{LK} \rapK  \big)
    \bigg) \nonumber \\
     &=& \sds \, \frac{\re{{P^{K \cup L}}}{x-x'} }{1 -   \re{J_K \cap J_L}{LK}\!^2 } 
     \NL \tim \bigg(  \re{J_K \cap J_L}{x-x'} \big(1 -   \re{J_K \cap J_L}{LK}\!^2 \big) -  \raK \rapK  - \re{J_L}{\aL} \re{J_L}{\apL}    \nonumber \\
&&  \quad \quad\quad\quad\quad \quad \quad \quad \quad + \; \re{J_K \cap J_L}{LK} \big( \raK \re{J_L}{\apL}  +  \re{J_L}{\aL} \rapK  \big)  \bigg) \label{eq_covLK1_para}
\ea
which is also explicitly invariant under the interchange of the two boundaries $ \mk \leftrightarrow \ml $.

The adjusted variance is obtained by setting $x=x'$ to get
\ba
\lefteqn{\vard{K \cup L}{f(x)}  } \nonumber \\
  &=& \sds \, \frac{1}{1 -   \re{J_K \cap J_L}{LK}\!^2 } 
  \NL \bigg( 1 -   \re{J_K \cap J_L}{LK}\!^2  -  \raK\!^2  - \re{J_L}{\aL}\!^2  
 + \; 2 \, \re{J_K \cap J_L}{LK} \raK \re{J_K}{\aL}   \bigg) \nonumber \\ \label{eq_varLK1_para}
\ea


\section{Proof for Multiple Intersecting Orthogonal Boundaries \label{Perp_proof} }

Here we prove Equations (\ref{ExKK}) and (\ref{CovKK}) by induction.

We first assume that the expressions hold for $ h-1 $ perpendicular boundaries, that is:
\small{
\ba
\ef{\KHm}{x} & =& \e{f(x)} + \sumihm \, (-1)^{i+1} \, \sumTHm \, \rKT \Delta f(\xKT) \\
\covDff{\KHm}{x}{x'} & = & \sds \rp{H-1}   \RHm \\
\RHm & = & \sumihz (-1)^i  \, \sumTHm  \rHmnT{x-x'}  \rT{\akHm}  \rT{\apkHm} \nonumber
\ea
}
\normalsize   
where $ H_{-1} = H-1 = \seq{1}{h-1} $  and $ J_{H-1} = \bigcup_{i \in H_{-1}} $.
We also assume that $ \simulator(x) $ is analytically solvable along $ \mk_1,\cdots,\mk_h $, permitting a large but finite number of evaluations to be performed along each boundary.  We can define an $ (m_i + 1) $-vector of boundary values to represent each boundary $ \mk_i $ as follows:
\be
K_i = (f(x^{K_i}), f(y_i^{(1)}),\cdots,f(y_i^{(m_i)})\ian{)}^T
\ee
which includes the projection of $ x^{K_i} $ of $ x $ onto $ \mk_i $.  To obtain an Equation analogous to Equation \eqref{eq_covfxK} we observe that:
\small
\ba
\lefteqn{\covDff{\KHm}{\xKh}{\zs}} 
\AEq \sds \rpHm{\xKh - \zs}   \R{\KHm}{\xKh}{\zs}
\AEq \sds  \rpH{x-\zs}  \sumihz (-1)^i \,  \sumTHm   \rHmnToh{x-\zs}    \rTnh{\zs - \zsKHm}
\ea
\ba
\lefteqn{\covDff{\KHm}{x}{\zs}}
\AEq \sds \rpHm{x-\zs}   \R{\KHm}{x}{\zs}
\AEq \sds   \rpH{x-\zs}  \rhnHm{x-\zs}
\NL   \sumihz (-1)^i \,   \sumTHm  \rHmnToh{x-\zs}  \rHmhnT{x-\zs} 
\NL \qq \qq \rTnh{\akHm}  \rTh{\akHm}  \rTnh{\zs - \zsKHm}
\AEq \sds \rpH{x-\zs}  \rhnHm{x-\zs} 
\NL \sumihz (-1)^i \,  \sumTHm   \rHmh{\akHm}  \rHmnToh{x-\zs}  
\NL \qq \qq \qq \qq \rTnh{\akHm}  \rTnh{\zs - \zsKHm}
\AEq  \rhnHm{x-\zs}   \rHmh{\akHm}   \covDff{\KHm}{\xKh}{\zs}
\AEq  \rh{\aKh}    \covDff{\KHm}{\xKh}{\zs} 
\ea
\normalsize
where we have used the fact that $ \rHmhnT{x-\zs} \rTh{\akHm}  =  \rHmh{\akHm} $.  Therefore:
\be
\covd{\KHm}{f(x)}{K_h}  =  \rh{\aKh}   \covd{\KHm}{f(\xKh)}{K_h}
\ee

\ian{Using the sequential Bayes linear expectation update Equation~(\ref{eq_BLmDK}), we} then have that: 
\small{
\ba
\lefteqn{ \ef{\KH}{x} } \nonumber\\
 &=& \ef{\KHm}{x}  + \covd{\KHm}{f(x)}{K_h} \vard{\KHm}{K_h}^{-1}(K_h- \ed{\KHm}{K_h})  \nonumber \\
&=&  \ef{\KHm}{x} +  \rh{\aKh}  (f(x^{K_h}) - \ed{\KHm}{f(x^{K_h})} ) \nonumber  \\ 
%
%
 &=&   \e{f(x)} +  \sumihm \, (-1)^{i+1} \, \sumTHm \, \rKT \Delta f(\xKT)   \nonumber \\ 
&& \quad  \;+\; \rh{\aKh}  \bigg(  f(x^{K_h}) - \e{f(x^{K_h})} \nonumber \\
&& \qq \qq \qq \qq \;-\; \sumihm \, (-1)^{i+1} \, \sumTHm \, \rTnh{\akT} \Df{\xKhKT}\bigg) \nonumber \\
&=&        \ef{}{x}   +   \rh{\akH}  \Df{\xKh}  
\NL +   \sumihm   \,  (-1)^{i+1} \, \sumTHm  \bigg(   \rT{\akT}  \Df{\xKT}  - \rToh{\aKhKT} \Df{\xKhKT} \bigg)   \nonumber \\
 &=&  \e{f(x)} \;+\; \sum_{i=1}^h \, (-1)^{i+1} \, \sumTH \, \rKT \Df{\xKT}   \label{eq_expKLgen}
\ea
}
\normalsize
and \ian{using the sequential Bayes linear covariance update, Equation~(\ref{eq_BLcDK}), we have} that:
\small
\ba
\lefteqn{ \covDff{\KH}{x}{x'} } 
\AEq  \covDff{\KHm}{x}{x'} \;-\; \covd{\KHm}{f(x)}{K_h} \vard{\KHm}{K_h}^{-1}\covd{\KHm}{K_h}{f(x')}  
%
%
\AEq \covDff{\KHm}{x}{x'}  \;-\; \rh{\aKh} \covd{\KHm}{f(x^{K_h})}{f(x')}  
\AEq  \covDff{\KHm}{x}{x'}  \;-\; \rh{\aKh} \covd{\KHm}{f(x^{K_h})}{f(\xpKh)}  \rh{\apKh} 
\AEq  \sds  \rp{H-1}  \R{\KHm}{x}{x'}  
\NL  - \rh{\aKh}  \sds  \rp{H}   \R{\KHm}{\xKh}{\xpKh}   \rh{\apKh}
\AEq  \sds  \rp{H}   \bigg(   \rhnHm{x-x'}  \R{\KHm}{x}{x'} 
\NL  \qq  \qq \qq  - \,  \rh{\aKh}  \rh{\apKh}    \R{\KHm}{\xKh}{\xpKh}   \bigg)
\AEq  \sds  \rp{H}   \bigg(   \rhnHm{x-x'}
\NL \sum_{i=0}^{h-1}  \, (-1)^i \, \sumTHm \, \rHmnT{x-x'}  \rT{\akHm} \rT{\apkHm}
\NL  \qq - \, \rh{\aKh}  \rh{\apKh} \sum_{i=0}^{h-1}  (-1)^i  \,   \sumTHm  \rHmnT{\xKh - \xpKh}  
\NL  \qq \qq \qq \tim \rT{\xKh - \xKhKHm}  \rT{\xpKh - \xpKhKHm}     \bigg)
\AEq  \sds  \rp{H}   
\NL \sum_{i=0}^{h-1} (-1)^i  \sumTHm  \bigg(  \rHmohnT{x-x'}  \rT{\akHm}  \rT{\apkHm} 
\NL \qq \qq - \,    \rHmnToh{x-x'}   \rTnh{\akHm}   \rTnh{\apkHm}  \rh{\aKh}  \rh{\apKh}    \bigg)
\AEq  \sds  \rp{H}   
\NL \sum_{i=0}^{h-1} (-1)^i  \sumTHm  \bigg(  \rHmohnT{x-x'}  \rT{\akH}  \rT{\apkH} 
\NL \qq  \qq  \qq \qq  \qq   - \, \rHmnToh{x-x'} \rToh{\aKh} \rToh{\apKh}  \bigg)
\AEq \sds \RH  \,  \rp{H}  \label{eq_covKLgen2}
\ea
\normalsize
Since the case for $ h=1 $ was derived in Section \ref{subsec:1KB:KBE}, this completes the proof. \begin{flushright} $ \Box $ \end{flushright}


\section{Proof for Multiple Parallel Boundaries \label{MultParB} }

Here we prove Equations (\ref{Epar}) and (\ref{Covpar}) by induction.

We begin by assuming that the expressions hold for $ h-1 $ parallel boundaries, that is:
\ba
\ef{\KHm}{x} & = & \e{f(x)} + \re{J_1}{a^{K_1}} \Delta f(x^{K_1}) \nonumber \\
&& \quad + \sum_{\gamma=2}^{h-1} \frac{ \Rg{\gmm}{\KGm}{x}{\Kg}}{ \Rg{\gmm}{\KGm}{\Kg}{\Kg}} \rgnGm{\aKg}  \bigg( \Df{\xKg} \nonumber \\
&& \qq  \,+\, \sumig \sumbG (-1)^{i+1}  \, \prodli \frac{ \Rg{\blm}{\KBlm}{\Kg}{\Kbl} }{ \Rg{\blm}{\KBlm}{\Kbl}{\Kbl} } \nonumber \\
&& \qq \quad \quad \quad \tim \rblnBlm{\Kblp \Kbl} \,\, \Df{\xKb} \bigg)  \\
\covDff{\KHm}{x}{x'} & = & \sds \rp{H-1} \Rg{h-1}{\KHm}{x}{x'}
\ea
We also assume that $ \simulator(x) $ is analytically solvable along $ \mk_1,...,\mk_h $, permitting a large but finite number of evaluations to be performed along each boundary.  We can define a $ (m_i + 1) $-vector of boundary values to represent each boundary $ \mk_i $ as follows:
\be
K_i = \big(f(x^{K_i}), f(y_i^{(1)}),...,f(y_i^{(m_i)})\big)^T
\ee
which includes the projection of $ x $ onto $ \mk_i $.  We first need to find an expression which relates $\covd{\KHm}{f(x)}{K_h}$ to $\covd{\KHm}{f(x^{K_h})}{K_h}$.  Noting that:
\ba 
\covd{\KHm}{f(x^{K_h})}{f(\ys_h)}  & = & \sds \re{P^{H-1}}{\xKh - \ys_h} \Rg{h-1}{\KHm}{K_h}{K_h}  \nonumber \\
& = & \sds \re{P^H}{x-\ys_h} \Rg{h-1}{\KHm}{K_h}{K_h} 
\ea
It follows that:
\ba 
\lefteqn{ \covd{\KHm}{f(x)}{f(\ys_h)} } \nonumber \\
& = & \sds \re{P^{H-1}}{x-\ys_h} \Rg{h-1}{\KHm}{x}{K_h}  \nonumber \\
& = & \rhnHm{\aKh} \frac{ \Rg{h-1}{\KHm}{x}{K_h} }{ \Rg{h-1}{\KHm}{K_h}{K_h}  }   \sds \re{P^H}{x-\ys_h} \Rg{h-1}{\KHm}{K_h}{K_h} \nonumber \\
& = & \frac{ \Rg{h-1}{\KHm}{x}{K_h} }{ \Rg{h-1}{\KHm}{K_h}{K_h}  } \rhnHm{\aKh}  \covd{\KHm}{f(x^{K_h})}{f(\ys_h)}
\ea
Therefore we have:
\be 
\covd{\KHm}{f(x)}{K_h}  
 =  \frac{ \Rg{h-1}{\KHm}{x}{K_h} }{ \Rg{h-1}{\KHm}{K_h}{K_h}  } \rhnHm{\aKh}  \covd{\KHm}{f(x^{K_h})}{K_h}
\ee
implying that we can again avoid explicit evaluation of the intractable $ \vard{\KHm}{K_h}^{-1} $ term.  Therefore the adjusted expectation can be calculated, using a sequential Bayes linear update, to be:
\footnotesize

\ba 
\lefteqn{ \ef{\KH}{x} } \nonumber \\
& = & \ef{\KHm}{x} \; + \; \covd{\KHm}{f(x)}{K_h} \vard{\KHm}{K_h}^{-1}(K_h- \ed{\KHm}{K_h})  \nonumber \\
& = & \ef{\KHm}{x} + \frac{ \Rg{h-1}{\KHm}{x}{K_h} }{ \Rg{h-1}{\KHm}{K_h}{K_h}  } \rhnHm{\aKh} \big( f(x^{K_h}) - \ed{\KHm}{f(x^{K_h})} \big) \nonumber \\
& = & \e{f(x)} + \re{J_1}{a^{K_1}} \Delta f(x^{K_1})\nonumber \\
&& \, \;+\; \sum_{\gamma=2}^{h-1} \frac{ \Rg{\gmm}{\KGm}{x}{\Kg}}{ \Rg{\gmm}{\KGm}{\Kg}{\Kg}} \rgnGm{\aKg} \bigg( \Df{\xKg} \nonumber \\ 
&& \quad + \, \sumig \sumbG (-1)^{i+1} \prodli \frac{ \Rg{\blm}{\KBlm}{\Kg}{\Kbl} }{ \Rg{\blm}{\KBlm}{\Kbl}{\Kbl} }  \rblnBlm{\Kblp \Kbl} \,\, \Df{\xKb} \bigg) \nonumber \\
&& \, \;+\; \frac{ \Rg{h-1}{\KHm}{x}{K_h} }{ \Rg{h-1}{\KHm}{K_h}{K_h}  } \rhnHm{\aKh} \bigg( f(x^{K_h}) - \e{f(x^{K_h})} - \re{J_1}{K_1K_h} \Delta f(x^{K_hK_1}) \nonumber \\ 
&& \quad \;-\; \sum_{\gamma=2}^{h-1} \frac{ \Rg{\gmm}{\KGm}{K_h}{\Kg} }{ \Rg{\gmm}{\KGm}{\Kg}{\Kg}} \rgnGm{K_h K_\gamma}   \bigg( \Delta f(x^{K_hK_\gamma}) 
\nonumber \\
&& \qq \, + \, \sumig \sumbG (-1)^{i+1}  \prodli \frac{ \Rg{\blm}{\KBlm}{\Kg}{\Kbl} }{ \Rg{\blm}{\KBlm}{\Kbl}{\Kbl} } \nonumber \\
&& \qq \qq \qq \qq \qq \qq  \qq \qq  \qq \quad \quad \quad \tim \rblnBlm{\Kblp \Kbl} \,\, \Df{x^{K_h K_b}} \bigg) \bigg) \nonumber \\
& = & \e{f(x)} + \re{J_1}{a^{K_1}} \Delta f(x^{K_1}) \nonumber \\
&& \, \;+\; \sum_{\gamma=1}^h \frac{ \Rg{\gmm}{\KGm}{x}{\Kg}}{ \Rg{\gmm}{\KGm}{\Kg}{\Kg}} \rgnGm{\aKg} \bigg( \Df{\xKg} \nonumber \\ 
&& \quad \;+\;  \sumig \sumbG (-1)^{i+1} \prodli \frac{ \Rg{\blm}{\KBlm}{\Kg}{\Kbl} }{ \Rg{\blm}{\KBlm}{\Kbl}{\Kbl} } \nonumber \\
&& \qq \qq \qq \qq \qq \qq \qq \qq \tim \rblnBlm{\Kblp \Kbl} \,\, \Df{\xKb} \bigg) 
\ea
\normalsize
Similarly, we also have that:
\ba 
\lefteqn{ \covDff{\KH}{x}{x'} } \nonumber \\
& = & \covDff{\KHm}{x}{x'} \covd{\KHm}{f(x)}{K_h} \vard{\KHm}{K_h}^{-1}\covd{\KHm}{K_h}{f(x')}  \nonumber \\
& = & \covDff{\KHm}{x}{x'} \nonumber \\
&& \quad \;-\; \frac{ \Rg{h-1}{\KHm}{x}{K_h} }{ \Rg{h-1}{\KHm}{K_h}{K_h}  } \rhnHm{\aKh} \covd{\KHm}{f(x^{K_h})}{f(x')} \nonumber \\
& = & \covDff{\KHm}{x}{x'} \nonumber \\
&& \quad \;-\; \frac{ \Rg{h-1}{\KHm}{x}{K_h} }{ \Rg{h-1}{\KHm}{K_h}{K_h}  } \rhnHm{\aKh} \nonumber \\
&& \qq \qq  \tim\covd{\KHm}{f(x^{K_h})}{f(x'^{K_h})} \rhnHm{\apKh} \frac{ \Rg{h-1}{\KHm}{a'}{K_h \KHm} }{ \Rg{h-1}{\KHm}{K_h}{K_h}  } \nonumber \\
& = & \sds \rp{H-1} \Rg{h-1}{\KHm}{x}{x'} \nonumber \\
&& \quad \;-\; \sds \rp{H} \frac{ \Rg{h-1}{\KHm}{x}{K_h} \Rg{h-1}{\KHm}{a'}{K_h \KHm} }{ \Rg{h-1}{\KHm}{K_h}{K_h} } \nonumber \\
&& \qq \qq \tim \rhnHm{\aKh} \rhnHm{\apKh} \nonumber \\
& = & \sds \rp{H}  \bigg( \Rg{h-1}{\KHm}{x}{x'} \rhnHm{x-x'} \nonumber \\
&& \quad \;-\; \frac{ \Rg{h-1}{\KHm}{x}{K_h} \Rg{h-1}{\KHm}{a'}{K_h \KHm} }{ \Rg{h-1}{\KHm}{K_h}{K_h} }  \rhnHm{\aKh} \rhnHm{\apKh}      \bigg) \nonumber \\ 
& = & \sds \rp{H} \Rg{h}{\KH}{x}{x'} 
\ea
Since the case for $ h=2 $ was derived in Section \ref{Mult_para_bound}, this completes the proof.  \begin{flushright} $ \Box $ \end{flushright}


\section{Derivation of Expressions \eqref{EgE} and \eqref{EgV} \label{Eg_App} }

Here we provide the derivation of Expressions \eqref{EgE} and \eqref{EgV}, presented as part of the example of Section \ref{TDE}.

Firstly, from Equations \eqref{Ex_two_para} and \eqref{Cov_two_para}, we have that the adjusted expectation and covariance of $ f(x), f(x') $ by $ \mk $ and $ \ml $ are given by:
\small
\ba
\lefteqn{\ed{K \cup L}{f(x)}} \AEq \e{f(x)} + \re{2,3}{a^K} \Df{\xK} + \frac{\R{2,3}{\aK}{LK}}{\R{2,3}{LK}{LK}} \Df{\xL} - \frac{\R{2,3}{\aK}{LK}}{\R{2,3}{LK}{LK}} \re{2,3}{LK} \Df{x^{LK}} \nonumber \\
\lefteqn{\covd{K \cup L}{f(x)}{f(x')}} \AEqN  \sds \re{1}{x-x'} \Rg{2}{K,L}{x}{x'}
\ea
\normalsize
To obtain an equation analogous to Equation \eqref{eq_covfxK}, we note that, for $ \ys \in \mm $:
\ba
\covd{K \cup L}{f(x^M)}{f(\ys)} & = & \sds \Rg{2}{K,L}{x^M}{\ys}  \\
\covd{K \cup L}{f(x)}{f(\ys)} & = & \sds \re{1}{x-\ys} \Rg{2}{K,L}{x}{\ys} \AEq
\sds \re{1}{a^M}  \Rg{2}{K,L}{x^M}{\ys} \AEq
\re{1}{a^M} \covDff{K \cup L}{x^M}{\ys}
\ea
Using the sequential Bayes linear update Equation \eqref{eq_BLmDK}, we then have that
\ba
\lefteqn{\ed{K \cup L \cup M}{f(x)}} 
\AEq \ed{K \cup L}{f(x)} + \re{1}{a^M}(f(x^M) - \ed{K \cup L}{f(x^M)})
\AEq \e{f(x)} + \re{2,3}{a^K} \Df{\xK} + \frac{\R{2,3}{\aK}{LK}}{\R{2,3}{LK}{LK}} \Df{\xL}
 - \frac{\R{2,3}{\aK}{LK}}{\R{2,3}{LK}{LK}} \re{2,3}{LK} \Df{x^{LK}}
 \NL + \, \re{1}{a^M} \bigg( f(x^M) - \bigg( \e{f(x^M)} + \re{2,3}{a^K} \Df{x^{MK}} 
 \NL \qq \qq \qq +  \, \frac{\R{2,3}{\aK}{LK}}{\R{2,3}{LK}{LK}} \Df{x^{ML}}
 - \frac{\R{2,3}{\aK}{LK}}{\R{2,3}{LK}{LK}} \re{2,3}{LK} \Df{x^{MLK}} \bigg) \bigg) 
\AEq  \ef{}{x} + \ro{\aM} \Df{\xM} + \rtt{\aK} \big( \Df{\xK} - \ro{\aM} \Df{x^{MK}} \big)
\NL \!\!\!\!\!\!\!\! + \, \frac{\Rtt{\aK}{LK}}{\Rtt{LK}{LK}}  \big( \Df{\xL} - \ro{\aM} \Df{x^{ML}} \big)
\NL \!\!\!\!\! - \, \frac{\Rtt{\aK}{LK}}{\Rtt{LK}{LK}} \rtt{LK} \big( \Df{\xLK} - \ro{\aM} \Df{x^{MLK}}  \big)
\ea
Similarly, using the sequential Bayes linear update Equation \eqref{eq_BLcDK} we have that
\ba
\covDff{K \cup L \cup M}{x}{x'} & = & \covDff{K \cup L}{x}{x'} - \re{1}{a^M} \covDff{K \cup L}{x^M}{x'^M} \re{1}{a'^M}
\AEq \sds \Rg{2}{K,L}{x}{x'} \re{1}{x-x'} - \re{1}{a^M} \sds \Rg{2}{K,L}{x}{x'} \re{1}{a'^M}
\AEq \sds \Rg{2}{K,L}{x}{x'} \R{1}{a^M}{a'^M}
\ea
so that:
\be
\vard{K \cup L \cup M}{f(x)} = \sds \Rg{2}{K,L}{x}{x} \R{1}{a^M}{a^M}
\ee


\section{Known Boundaries and Black Box Emulation Packages \label{KBBB} }

In this section, we expand on the issues alluded to in Section \ref{subsec:UFME} that users of black box emulators might face when trying to implement Equations (\ref{eq_BLmDK})-(\ref{eq_BLcDK}), or analogous sequential update equations, for additional model runs $ D $ and multiple known boundaries directly.  As discussed in \cite{KBECCM} \ian{and implied by Equations~(\ref{eq_BLmDK})-(\ref{eq_BLcDK})}, given a single known boundary $ \mk $, a sufficient set of model evaluations for the full joint update by $K \cup D$ is composed of $ D $, $ D^K = ( f(x^{(1)K}),...,f(x^{(n)K}) ) $ and $ f(x^K) $; a total of $2n+1$. This has ramifications for users of black box Gaussian process emulation packages (such as
BACCO \citep{Hankin:2005aa} or GPfit \citep{JSSv064i12} in R, or \cite{gpy2014} in Python), 
which may not be easily recoded to use the analytic emulation formulae of Equations~(\ref{eq_EK1}) and (\ref{eq_VK1}). Such a user has to add the 
extra $(n+1)$ trivial evaluations $ \{ D^K, f(x^K) \} $ to their usual set of $n$ standard evaluations $ D $ to give $ D^* = \{ D, D^K, f(x^K) \} $, and \ian{then} their black box Gaussian process package will produce results that 
precisely match Equations~(\ref{eq_BLmDK})-(\ref{eq_BLcDK}). This will, however, require inverting a matrix of square size $2n+1 $ as a result of essentially using Equations (\ref{BLE1})-(\ref{BLE3_univariate}) with $ D $ replaced by $ D^* $,
and hence be slower than directly using the above analytic results, which only require inverting a matrix of square size $ n$, corresponding to the points in $ D $.

This reduction in computational efficiency may particularly cause issues for users of black box emulation packages if the sequential update, given by Equations (\ref{eq_BLmDK})-(\ref{eq_BLcDK}), is required for a large batch of $ n' $ points, since each point will require a matrix inversion \ian{corresponding to its own $D^*$}, as discussed above.  These emulation calculations can be made more efficient by emulating the $ n' $ model runs in $ N' $ batches $ \{ B_1,...,B_{N'} \} $, where we define a generic batch, of size $ n_B $,  to be $ B = (f(x_B^{(1)}),...,f(x_B^{(n_B)}))^T $.  Even in this case, each batch requires the black box emulation package to invert a matrix of square size $ |D^*| = 2n+n'_B $ (where now $ D^* = \{ D, D^K, B^K \}  $, with $ B^K = (f(x_B^{(1)K}),...,f(x_B^{(n_B)K}))^T$) in order to incorporate knowledge of boundary $ \mk $.  Careful choice of $ n'_B $ will improve emulator efficiency, however, this calculation may still be infeasible if the size of $ n $ and/or $ n' $ is too large.  In comparison, using the above analytic results 
\ian{(Equations~(\ref{eq_BLmDK})-(\ref{eq_BLcDK}) combined with~(\ref{eq_EK1}), (\ref{eq_covK2}), and (\ref{eq_VK1}))}
only requires inversion of a single $ n \times n $ matrix, regardless of the size of $ n' $.

Furthermore, the issues discussed above are exacerbated for the case of multiple known boundaries.  For the case of $ h $ perpendicular boundaries $ \mk_H = \seqsub{\mk}{1}{h} $ (as discussed in Section \ref{ssec_mult_orth_bound}), the model runs sufficient for calculating $ \ed{\KH \cup D}{B} $ and $ \vard{\KH \cup D}{B} $ are $ D^* = \{ D, \bigcup_{T \subseteq H} D^{K_A}, \bigcup_{T \subseteq H} B^{K_A}  \} $.  As a result, a black box approach requires a matrix inversion of square size $ |D^*| = 2^hn + (2^h -1)n_B $.  In the case of $ h $ parallel boundaries (Section \ref{Mult_para_bound}), sufficient information for calculating $ \ed{\KH \cup D}{B} $ and $ \vard{\KH \cup D}{B} $ is $ D^* = \{ D, D^{K_1},...,D^{K_h}, B^{K_1},...,B^{K_h} \} $, requiring a matrix inversion of square size $ |D^*| = (h+1)n + hn_B $.  Either way, if $ h $ is not small and/or either of $ n, n_B $ are large, a black box emulator has to deal with substantial matrix inversions.  On the other hand, \ian{encoding the results analytically still only involves a matrix inversion of size $n$}, hence may be \ian{significantly advantageous}.


\section{Multivariate Emulation \label{ME} }

In this section, we assume that we have a $ q $-variate computer model $ f(x) \in \real^q $.  We discuss the generalisation of the previous results to multivariate emulators with a separable correlation structure (see, for example, \cite{LEMDF} and \cite{BECMODCM}), that is, one of the form:
\begin{equation}
\cov{f(x)}{f(x')} = c(x - x') \, \Sig = \prod_{j=1}^p r_j(x_j - x_j') \, \Sig  \label{Mult_frame}
\end{equation}
The structure of Equation \eqref{Mult_frame} is analogous to that presented in Equation \eqref{eq_cor_struc}, with $ \sigma^2 $ replaced by $ \Sig \in \real^{q \times q} $, a $ q \times q $ covariance matrix between the output components with $ \Sig_{ii'} $ representing the covariance between output components $ v $ and $ v' $ evaluated at any inputs $ x $ and $ x' $.
If the behaviour of each output component of the computer model is known along the corresponding boundaries, then the results for expectation are as presented in the previous sections, and the results for covariance are similar to those presented, with the only difference being the replacement of $ \sds $ by covariance matrix $ \Sig $ in the appropriate places.  This follows since the previous results in this article have directly comparable results in terms of the correlation between two inputs $ x $ and $ x' $ as they do for covariance (with the only difference being a scaling by a constant $ \sds $).  As an example, we present here the calculations for the single boundary case.

As before, we extend the collection of boundary evaluations $ K $ to be the $ (m+1)q $ column vector:
\be
K=(f(x^K),f(y^{(1)}),\dots,f(y^{(m)}))^T \label{multi_opt_K}
\ee
The trivial identity used to obtain Equation \eqref{eq_covvar} still holds but is now given more generally by:
\begin{align}
\boldsymbol{I}_{(m+1)q}  &=\;\var{K} \var{K}^{-1} \\
	 &=\; \begin{pmatrix} 
	\covb{f(x^K)}{K} \\
	\covb{f(y^{(1)})}{K} \\
	\vdots \\
	\covb{f(y^{(m)})}{K} 
	\end{pmatrix}  \var{K}^{-1}.   
\end{align}
so that
\begin{equation}
\cov{f(x^K)}{K} \var{K}^{-1} = ( \boldsymbol{I}_{q} \,\,  \boldsymbol{0}_{q \times mq} )
\end{equation}

Corresponding to Equation (\ref{eq_shortcov2}) we have:
\ba
\covb{f(x)}{f(x^K)} &=& \Sig \, \re{P}{x-x^K} \;=\; \Sig \, \raK \nonumber  \\
\ea 
Furthermore, we then have, corresponding to Equation (\ref{eq_prod_ra_xy}):
\ba
\covb{f(x)}{f(\ys)} &=& \Sig \, \re{P}{x-\ys}  \nonumber \\
&=& \Sig \, \raK \re{{P^K}}{x-\ys} \nonumber \\
&=& \raK \, \covb{f(x^K)}{f(\ys)} 
\ea 
and Equation (\ref{eq_covfxKpre}) still holds.

The Bayes linear update equations for $ f(x) $ with respect to $ K $ now give:
\ba
\ed{K}{f(x)} & = & \e{f(x)} + \cov{f(x)}{K} \var{K}^{-1}(K- \e{K}) \nonumber \\
&=& \e{f(x)} + \raK ( \boldsymbol{I}_{q} \,  \boldsymbol{0}_{q \times mq} )  (K- \e{K}) \nonumber \\
&=&  \e{f(x)} + \raK \Delta f(x^K) 
\ea
\ba
\covd{K}{f(x)}{f(x')} &=& \cov{f(x)}{f(x')} - \cov{f(x)}{K} \var{K}^{-1} \cov{K}{f(x')} \nonumber \\
&=&  \cov{f(x)}{f(x')} -  \raK ( \boldsymbol{I}_{q} \,  \boldsymbol{0}_{q \times mq} ) \cov{K}{f(x')}  \nonumber \\
&=& \cov{f(x)}{f(x')} -  \raK  \cov{f(x^K)}{f(x'^K)} \rapK   \nonumber  \\
&=& \Sig \, \RaK \, \re{{P^K}}{x-x'}
\ea

Although the above result is nice, it is likely that boundary behaviour may only be known for some (and not all) output components.  In this case, one may wish to use the multivariate correlation structure to update one's beliefs about all output components given knowledge of the behaviour of one component.  Such calculations can still be performed analytically for certain combinations of boundaries and output components.  As an example, we will present the calculations required to update the expectation of, and the covariance between, two output components, given that the behaviour on a single boundary $ \mk $ is known for a third component.  

Corresponding to Equation (\ref{eq_shortcov2}) we have:
\ba
\covb{f_2(x)}{f_1(x^K)} &=& \Sig_{21} \re{P}{x-x^K} \;=\; \Sig_{21} \raK \nonumber  \\
& = & \raK \cov{f_2(x^K)}{f_1(x^K)}
\ea 
where $ \Sig_{vv'} $ denotes the covariance between output components $ v $ and $ v' $.
Furthermore, we then have, corresponding to Equation (\ref{eq_prod_ra_xy}):
\ba
\covb{f_2(x)}{f_1(\ys)} &=& \Sig_{21} \re{P}{x-\ys}  \nonumber \\
&=& \Sig_{21} \raK \re{{P^K}}{x-\ys} \nonumber \\
&=& \raK \, \covb{f_2(x^K)}{f_1(\ys)} 
\ea 
and then, corresponding to Equation (\ref{eq_covfxKpre}), we have:
\ba
\cov{f_2(x)}{K^1} & = & \raK \cov{f_2(x^K)}{K^1} 
\ea
where the notation $ K^v = (f_v(x^K), f_v(y^{(1)}),\cdots,f_v(y^{(m)}))^T $ represents evaluation of model output component $ v $ at a large set of points along boundary $ \mk $.
We then have that:
\ba
\cov{f_2(x)}{K^1} \var{K^1}^{-1} & = &  \raK \cov{f_2(x^K)}{K^1} \var{K^1}  \nonumber \\
& = & \frac{\Sig_{21}}{\Sig_{11}} \raK (1,0,...,0)
\ea
So that the Bayes linear update equations result in:
\ba
\ed{K^1}{f_2(x)} & = & \e{f_2(x)} + \cov{f_2(x)}{K^1} \var{K^1}^{-1}(K^1- \e{K^1}) \nonumber  \\
& = & \e{f_2(x)} + \frac{\Sig_{21}}{\Sig_{11}} \, r_{J_K}(\aK) \Delta_1 f(x^K) 
\ea
where $ \Delta_1 f(x^K) = f_1(x^K) - \e{f_1(x^K)} $, and:
\ba
\lefteqn{\covd{K^1}{f_2(x)}{f_3(x')}} \nonumber \\
 & = & \cov{f_2(x)}{f_3(x')} - \cov{f_2(x)}{K^1} \var{K^1} \cov{K^1}{f_3(x')} \nonumber              \\
& = & \Sig_{23} \raaK- \frac{\Sig_{21}}{\Sig_{11}} \raK \cov{f_1(x^K)}{f_3(x)}  \nonumber \\
& = & \big( \Sig_{23} \,  r_{J_K}(\aK-\apK) - \frac{\Sig_{21}\Sig_{31}}{\Sig_{11}} \, r_{J_K}(\aK) r_{J_K}(\apK) \big) r_{{P^K}}(x-x') \label{cov_eq_MO}
\ea
Although this update was relatively straightforward, our updated beliefs about the behaviour of output component 2 based on the known behaviour along boundary $ \mk $ of output component 1 no longer have a product correlation structure, or indeed a separable variance structure, as can be seen by looking at the corresponding variance equation to Equation (\ref{cov_eq_MO}), namely:
\ba
\hspace{-0.2cm} \vard{K^1}{f_2(x)} & = & \big( \Sig_{22} \,  r_{J_K}(\aK-\apK) - \frac{\Sig_{21}^2}{\Sig_{11}} \, r_{J_K}(\aK) r_{J_K}(\apK) \big) r_{{P^K}}(x-x') \label{var_eq_MO}
\ea
Hence, updating our beliefs by further boundaries may not be possible analytically.  The natural question to ask is therefore: for which combinations of boundaries can an analytical update be achieved?  The answer to this question follows naturally from the proposition in Section \ref{PerpSetsPar}.  Due to the separable product correlation structure across the input parameters, we can view the output component indicator as an additional parameter to a scalar-output computer model.  In other words, we can view the parameters as being: $ x_1,...,x_p,x_{opt} $, where $ x_{opt} $ indicates for which output component the computer model is being evaluated.  Following this, the answer to the question is as given in Section \ref{PerpSetsPar}.


\section{Arabidopsis Example: Further Discussion \label{App:AMS} }

Here we present extended detail about the Arabidopsis model of Section \ref{AMAM}.  Table \ref{DE} shows the full Arabidopsis model, represented as a set of 18 differential equations.  This full model takes an input vector of 45 rate parameters $ (k_1, k_{1a}, k_2, ...) $ and produces an output vector of 18 chemical concentrations $ ([Auxin], [X], [PLSp], ... ) $.  Table \ref{RR} shows the ranges over which we allowed each parameter of the Arabidopsis model to vary, this describing the input space of interest $ \mx $.

\begin{table}
\scriptsize
\begin{align*}
\frac{d[Auxin]}{dt}  \;\;=\;\; &  \frac{k_{1a}}{\displaystyle 1 + \frac{[X]}{k_1}} + k_2 + k_{2a}   
\frac{[ET]}{\displaystyle 1 + \frac{[CK]}{k_{2b}}} \frac{[PLSp]}{k_{2c} + [PLSp]}   
& \frac{d[Re]}{dt}  \;\;=\;\; & k_{11}[Re^\ast][ET] - (k_{10} + k_{10a}[PLSp])[Re]  
\\ 
&  + \frac{V_{IAA}[IAA]}{Km_{IAA} + [IAA]}   
& \frac{d[Re^\ast]}{dt} \;\;=\;\; & -k_{11}[Re^\ast][ET] + (k_{10} + k_{10a}[PLSp])[Re]   \\
&  - \left( k_3 + \frac{k_{3a}[PIN1pm]}{k3auxin + [Auxin]} \right) [Auxin]    
&\frac{d[CTR1]}{dt}  \;\;=\;\; & -k_{14}[Re^\ast][CTR1] + k_{15}[CTR1^\ast]   
\\
\frac{d[X]}{dt}  \;\;=\;\; & k_{16} - k_{16a}[CTR1^\ast] - k_{17}[X]  
&\frac{d[CTR1^\ast]}{dt}  \;\;=\;\; & k_{14}[Re^\ast][CTR1] - k_{15}[CTR1^\ast]   \\
\frac{d[PLSp]}{dt}  \;\;=\;\; & k_8[PLSm] - k_9[PLSp]   
&\frac{d[PIN1m]}{dt}  \;\;=\;\; & \frac{k_{20a}}{k_{20b} + [CK]} [X] \frac{[Auxin]}{k_{20c} + [Auxin]}     \\ 
\frac{d[Ra]}{dt}  \;\;=\;\; & -k_4 [Auxin] [Ra] + k_5 [Ra^\ast]   
& &- k_{1_v21}[PIN1m]  \\
\frac{d[Ra^\ast]}{dt}  \;\;=\;\; & k_4 [Auxin] [Ra] - k_5 [Ra^\ast]  
&\frac{d[PIN1pi]}{dt}  \;\;=\;\; & k_{22a}[PIN1m] - k_{1_v23}[PIN1pi]   \\ 
\frac{d[CK]}{dt}  \;\;=\;\; & \frac{k_{18a}}{\displaystyle 1 + \frac{[Auxin]}{k_{18}}} - k_{19} [CK]   
& & - k_{1_v24}[PIN1pi] + \frac{k_{25a}[PIN1pm]}{\displaystyle 1 + \frac{[Auxin]}{k_{25b}}}   \\
& + \frac{V_{CK}[cytokinin]}{Km_{CK} + [cytokinin]}   
&\frac{d[PIN1pm]}{dt}  \;\;=\;\; & k_{1_v24}[PIN1pi] - \frac{k_{25a}[PIN1pm]}{\displaystyle 1 + \frac{[Auxin]}{k_{25b}}}   \\
\frac{d[ET]}{dt}  \;\;=\;\; & k_{12} + k_{12a}[Auxin][CK] - k_{13}[ET]   
&\frac{d[IAA]}{dt}  \;\;=\;\; & 0   \\
& + \frac{V_{ACC}[ACC]}{Km_{ACC} + [ACC]}   
&\frac{d[cytokinin]}{dt}  \;\;=\;\; & 0   \\
\frac{d[PLSm]}{dt}  \;\;=\;\; & \frac{k_6[Ra^\ast]}{\displaystyle 1 + \frac{[ET]}{k_{6a}}} - k_7[PLSm]   
&\frac{d[ACC]}{dt}  \;\;=\;\; & 0   
\end{align*}
\caption[Arabdiopsis Equations]{\small{Arabidopsis model differential equations.} \label{DE}}
\end{table}

Note that we do not explore different values of seven of the inputs. In particular, we fix each of $ V_{IAA} $, $ V_{CK} $ and $ V_{ACC} $ at 0 (note that this also effectively removes $ Km_{IAA}, Km_{CK} $ and $Km_{ACC} $).  These three inputs represent feeding the Arabidopsis plant with the hormones auxin, cytokinin and ethylene, \ivm{while} exploring the hormonal behaviour of the roots of the plant without feeding is also of substantial interest to biologists.  In addition, it is necessary to impose a further constraint that $ k_{16}/k_{16a} = 0.3 $, as presented in \cite{MEAHCA} and suggested by the results of \cite{BUCSBM}, which ensures that the term $ k_{16} - k_{16a}[CTR1^\ast] $ in the $ d[X]/dt $ equation is non-negative.  This constraint effectively removes another parameter, hence $ k_{16} $ in the differential equations of Table \ref{DE} above is calculated according to this constraint from the value of $ k_{16a} $.  The dimension of the model's input parameter space is therefore effectively 38.  
\begin{table}[h]
\spacingset{1}
	\center
	\begin{footnotesize}
	\begin{tabular}{|c|c|c|} \hline
		Input / Input Ratio  & Minimum & Maximum  \\ \hline
		$ k_1 $ &  0.1 & 4  \\
		$ k_{1a} $  & 0.1 & 4  \\ 
		$ k_2 $ & 0.02 & 0.8 \\
		$ k_{2a} $  & 0.28 & 11.2   \\
		$ k_{2b} $ & 0.1  & 4  \\
		$ k_{2c} $ &  0.001 & 0.04 \\
		$ k_3 $ & 0.2 & 8 \\
		$ k_{3a} $ & 0.045 & 1.8  \\
		$ k_{3auxin} $ & 1 & 40  \\
		$ k_{1_vauxin} $ & 0.1 & 4  \\
		$ k_4 $ &0.1 & 4 \\
		$ k_5 $ & 0.03 &  1.2  \\	 
		$ k_{6a} $ & 0.02 & 0.8  \\
		$ k_7 $  & 0.1 & 4  \\
		$ k_8 $ & 0.1 & 1  \\
		$ k_9 $ & 0.1 & 1 \\
		$ k_{10} $ & 0.00003 & 0.0012 \\
		$ k_{10a} $ & 0.5 & 20  \\
		$ k_{11} $ & 0.5 & 20  \\
		$ k_{12} $ & 0.01 & 0.4 \\
		$ k_{12a} $  & 0.01 & 0.4  \\
		$ k_{13} $  & 0.1 & 1 \\
		$ k_{14} $ &0.3 & 12\\
		$ k_{15} $ & 0.0085  & 0.34  \\
		$ k_{16a} $ & 0.1 & 4 \\
		$ k_{17} $ & 0.01 & 0.4  \\
		$ k_{18} $  & 0.01 & 0.4  \\
		$ k_{18a} $ & 0.1  & 4 \\
		$ k_{19} $ & 0.1 & 4 \\
		$ k_{20a} $ & 0.08  & 3.2  \\
		$ k_{20b} $ & 0.1 & 4  \\
		$ k_{20c} $ & 0.03  & 1.2  \\
		$ k_{1_v21} $ & 0.1 & 4 \\
		$ k_{22a} $ & 0.1 & 1  \\
		$ k_{1_v23} $ & 0.075 & 3 \\
		$ k_{1_v24} $ & 1 & 40 \\
		$ k_{25a} $ &  0.1 & 4  \\
		$ k_{25b} $ & 0.1 & 4   \\ \hline
	\end{tabular}
	\end{footnotesize}
	\caption[Input Ranges for Arabidopsis Model]{\small{Input parameter ranges (which underwent a square root transformation and were scaled to $ [-1,1] $ for analysis), that make up the input parameter space of interest $ \mx $.} \label{RR}}
\end{table}

Table \ref{IOV} shows the initial conditions of the output components of the Arabidopsis model.

\begin{table}[ht] 
\spacingset{1}
	\center
	\begin{tabular}{cc|cc}
		Output Component & Initial Condition & Output Component & Initial Condition \\ \hline 
		$ [Auxin] $ & 0.1 & $ [Re\ast] $ & 0.3 \\
		 $ [X] $ & 0.1 & $ [CTR1] $ & 0 \\ 
		 $ [PLSp] $ & 0.1 & $ [CTR1\ast] $ & 0.3 \\ 
		 $ [Ra] $ & 0 & $ [PIN1m] $ & 0 \\ 
		 $ [Ra\ast] $ & 1 & $ [PIN1pi] $ & 0 \\ 
		 $ [CK] $ & 0.1 & $ [PIN1pm] $ & 0 \\
		 $ [ET] $ & 0.1 & $ [IAA] $ & 0  \\ 
		 $ [PLSm] $ & 0.1 & $ [cytokinin] $ & 0 \\ 
		 $ [Re] $ & 0 & $ [ACC] $ & 0  \\
	\end{tabular}
	\caption[Arabidopsis Model Output Components]{\small{The list of 18 output components to the model of \cite{IPPHCARD}, along with their initial conditions.  See \cite{MEAHCA} and \cite{IPPHCARD} for details.}  \label{IOV}}
\end{table}

Table \ref{tab:RMSE} shows Root Mean Squared Errors (RMSE) for the 500 diagnostic points for each of the emulators constructed in Section \ref{SEPS}:
\begin{equation}
\sqrt{\frac{1}{500} \sum_{w=1}^{500} \big( f(x^{(w)}) - \mu(x^{(w)}) \big)^2 }
\end{equation}
These quantities are a measure of how accurate the emulator predictions are for the diagnostic points on average.  The most accurate predictions are obtained when $ \theta = 10 $ and there are no known boundaries, however, as discussed in relation to Table \ref{tab:SumVarMASPE}, this case results in an invalid emulator as a result of the assessment of uncertainty being too low.  RMSEs for $ \theta = 6 $ with 0 or 1 known boundary, or $ \theta = 3 $ with 2 known boundaries are then most accurate, each with a RMSE less than 0.16, however, the MASPEs when $ \theta = 6 $ are also too large.  It should be noted that the invalid diagnostics when known boundaries are included should not result in the exclusion of the known boundaries, but rather a reconsideration of our prior belief structure.  A suitable belief structure will result in valid diagnostics, even with the inclusion of known boundaries.  For the current application, perhaps the assumption of stationarity across the input space isn't appropriate.  In conclusion, inclusion of 2 known boundaries and 1000 training points provides the most information, and results in a valid emulator with greatly increased accuracy.

\begin{table}
\spacingset{1}
\footnotesize
\centering
\begin{tabular}{|c|ccccc|}
  \hline
$ \theta $ & 0.1 & 1 & 3 & 6 & 10 \\ 
  \hline
0TP, 0KB & 0.690 & 0.690 & 0.690 & 0.690 & 0.690 \\ 
  0TP, 1KB & 0.690 & 0.686 & 0.418 & 0.472 & 0.570 \\ 
  0TP, 2KB & 0.690 & 0.530 & 0.359 & 0.552 & 0.610 \\ 
  200TP, 0KB & 0.690 & 0.690 & 0.378 & 0.196 & 0.188 \\ 
  200TP, 1KB & 0.690 & 0.686 & 0.265 & 0.217 & 0.250 \\ 
  200TP, 2KB & 0.690 & 0.530 & 0.198 & 0.260 & 0.352 \\ 
  500TP, 0KB & 0.690 & 0.690 & 0.266 & 0.168 & 0.170 \\ 
  500TP, 1KB & 0.690 & 0.686 & 0.205 & 0.180 & 0.205 \\ 
  500TP, 2KB & 0.690 & 0.530 & 0.177 & 0.229 & 0.296 \\ 
  1000TP, 0KB & 0.690 & 0.690 & 0.223 & 0.142 & 0.136 \\ 
  1000TP, 1KB & 0.690 & 0.686 & 0.177 & 0.153 & 0.165 \\ 
  1000TP, 2KB & 0.690 & 0.530 & 0.159 & 0.210 & 0.265 \\ 
   \hline
\end{tabular}
\caption{\small{Root Mean Squared Errors for the set of diagnostic points for different values of common $ \theta $ and numbers of known boundaries (KB) and training points (TP) in the bulk of the input space.} \label{tab:RMSE}}
\end{table}

\printnomenclature

\end{document}